\documentclass[
  aps,
  prx,
  reprint,
  preprintnumbers,
  superscriptaddress,
  longbibliography,
  floatfix,
  amsmath,
  amssymb
]{revtex4-2}

\usepackage{graphicx}
\usepackage{fix-cm}
\usepackage{tikz}
\usetikzlibrary{arrows.meta,decorations.markings}
\usepackage{bm}
\usepackage{xcolor}
\usepackage{needspace}
\usepackage{hyperref}
\usepackage[normalem]{ulem}
\usepackage{comment}
\hypersetup{hidelinks}

\begin{document}

\preprint{USTC-ICTS/PCFT-26-81}

\title{CANON: Agent-Assisted Construction of Canonical Differential Equations for Feynman Integrals}
\author{Dong-Shan Jian}
\email{dsjian@stu.pku.edu.cn}
\affiliation{School of Physics, Peking University, Beijing 100871, China}

\author{Yuanche Liu}
\email{liuyuanche@mail.ustc.edu.cn}
\affiliation{Interdisciplinary Center for Theoretical Study, University of Science and Technology of China, Hefei, Anhui 230026, China}

\author{Yan-Qing Ma}
\email{yqma@pku.edu.cn}
\affiliation{School of Physics and State Key Laboratory of Nuclear Physics and Technology, Peking University, Beijing 100871, China}
\affiliation{Center for High Energy Physics, Peking University, Beijing 100871, China}

\author{Bo-Xuan~Shi}
\email{shibx@pku.edu.cn}
\affiliation{Center for High Energy Physics, Peking University, Beijing 100871, China}

\author{Yu Wu}
\email{wy626@mail.ustc.edu.cn}
\affiliation{LAPTh, Universit\'e Savoie Mont Blanc, CNRS, B.P.~110, F-74941 Annecy-le-Vieux, France}

\author{Zihao Wu}
\email{wuzihao@mail.ustc.edu.cn}
\affiliation{State Key Laboratory of Nuclear Physics and Technology, Institute of Quantum Matter, South China Normal University, Guangzhou 510006, China}
\affiliation{Guangdong Basic Research Center of Excellence for Structure and Fundamental Interactions of Matter, Guangdong Provincial Key Laboratory of Nuclear Science, Guangzhou 510006, China}

\author{Yang Zhang}
\email{yzhphy@ustc.edu.cn}
\affiliation{Interdisciplinary Center for Theoretical Study, University of Science and Technology of China, Hefei, Anhui 230026, China}
\affiliation{Center for High Energy Physics, Peking University, Beijing 100871, China}
\affiliation{Peng Huanwu Center for Fundamental Theory, Hefei, Anhui 230026, China}

\begin{abstract}
    The construction of canonical basis is an important procedure in the calculation of multi-loop Feynman integrals via the canonical differential equation methods. Currently, there are various different methods for this purpose. Among them, no single method is universally applicable. Thus, obtaining a valid basis often requires repeated attempts and verification. In this paper, we present CANON, an agent-assisted framework designed to make this search-and-verification process more convenient and more efficient. In this workflow, the agent selects methods and calls the corresponding tools to revise the basis proposal or the associated differential-equation system. Independent programs then verify whether the agent's result is valid and ask the agent to retry if verification fails. This framework contains a specific state-management tool to supports long-range basis searching works which requires multiple sub-agents. Its modular architecture enables future extensions to the framework itself. In this paper, we demonstrate this framework by applying it to some two-loop and three-loop integral families.
\end{abstract}

\maketitle
\raggedbottom

\section{Introduction}
\label{sec:introduction}

Precision measurements at the LHC serve as one of the key ways to test the Standard Model and probe
new physics. Achieving these goals also requires theoretical predictions with comparably small
uncertainties~\cite{Azzi:2019yne,Heinrich:2020ybq,Huss:2025nlt}.
Higher-order corrections to these predictions require multi-loop Feynman integrals. 
In the recent years, advances in multiloop techniques have enabled a wide range of
high-order calculations in recent years. These progresses include NNLO photon-plus-two-jet
predictions~\cite{Badger:2023mgf}, N$^{3}$LO $W$
distributions and single-inclusive hadron coefficient
functions in $e^+e^-$ annihilation~\cite{Chen:2022lwc,He:2025hin},
full-color two-loop five-parton amplitudes~\cite{Agarwal:2023suw},
and leading-color double-virtual contributions to $t\bar t+\mathrm{jet}$
production~\cite{Badger:2025ljy}. Massive examples include
two-loop electroweak corrections to gluon-fusion Higgs-pair
production~\cite{Bi:2023bnq} and two-loop heavy-quark-mediated
diphoton amplitudes~\cite{Becchetti:2025rrz}.

For processes concerning higher-order corrections, or with a larger number of particles and mass scales, which are currently of interest, the computation of multi-loop integrals remains a bottleneck. This is because the complexity of multi-loop computations grows rapidly as the number of loops, particles, and scales increases~\cite{Weinzierl:2022eaz,Abreu:2022mfk}. Consequently, multi-loop Feynman integral computation is still one of the major topics in high energy physics.

To compute Feynman integrals, one usually reduces the required integrals (including tensor reduction and integration-by-parts (IBP) reduction) to express them in terms of a finite set of master integrals (MIs)~\cite{Ossola:2006us,Ossola:2007ax,Passarino:1978jh,Tkachov:1981wb,Chetyrkin:1981qh,Laporta:2000dsw}.
The master integrals can be evaluated using various methods including the differential equation method~\cite{Kotikov:1990kg,Remiddi:1997ny,Gehrmann:1999as}. 
In this method, one usually tries to find a canonical basis, which can be obtained by a transformation of the original master integrals. This transformation make the dimensional regulator $\varepsilon$ factor out of the
differential-equation matrices, although the basis transformation can
itself depend on the regulator. These matrices, called canonical differential equation (CDE) matrices, makes it possible to organize the regulator expansion of the integrals in the basis as iterated integrals over an
alphabet~\cite{Henn:2013pwa,Henn:2014qga}, which is very useful especially in the polylogarithmic setting.

Suitably normalized integrals of uniform transcendental weight (UT)
are natural candidates for a canonical basis. Constructing such a basis, as well as its corresponding CDE, is one of the central bottlenecks of the CDE method. In practice, this construction
often requires combining many different methods, from the selection of
integral representatives to transformations of the differential equations
and the completion of lower-sector couplings~\cite{Gehrmann:2014bfa,Liu:2024ont}.

There are various integral-based constructions. They include the usage of logarithmic singularities and
constant leading singularities to select numerators and
normalizations~\cite{Arkani-Hamed:2010pyv,Arkani-Hamed:2014via,Bern:2014kca,Henn:2020lye}.
Logarithmic representations can also generate differential equations
directly~\cite{Herrmann:2019upk}.
Baikov and loop-by-loop representations support such searches in
multi-scale problems~\cite{Baikov:1996cd,Frellesvig:2017aai,Dlapa:2021qsl,He:2022ctv},
while dimensional leading singularities and Gram numerators capture
contributions invisible to four-dimensional
residues~\cite{Chicherin:2018old,Henn:2024ngj}.
Dimensional shifts can also yield quasi-finite representatives with
improved convergence~\cite{Tarasov:1996br,vonManteuffel:2014qoa}; quasi-finiteness
alone does not imply uniform transcendental weight, although finite
and UT bases can be related~\cite{Schabinger:2018dyi}.
Intersection numbers provide projections onto integral
bases~\cite{Mastrolia:2018uzb,Chen:2020uyk,Chen:2022lzr}.
Intersection theory and Schubert geometry also provide information
about alphabets, connections and basis
transformations~\cite{Chen:2023kgw,He:2023umf,Duhr:2024xsy,Duhr:2025xyy}.

Another direction consists of many direct differential-equation methods. These methods instead transform an existing
system. Lee's algorithm uses Fuchsian reduction and residue
normalization to factor out the regulator~\cite{Lee:2014ioa,Lee:2017oca};
\texttt{epsilon}, Fuchsia and Libra provide implementations and
transformation tools~\cite{Prausa:2017ltv,Gituliar:2017vzm,Lee:2020zfb}.
Meyer's algorithm and CANONICA solve rational-function ansatzes for the
transformation, including in several kinematic
variables~\cite{Meyer:2016slj,Meyer:2017joq}.
For suitable regulator-linear systems, Magnus expansions construct a
transformation that removes the regulator-independent part of the
connection~\cite{Argeri:2014qva}.
Building on an earlier seed-based construction~\cite{Hoschele:2014qsa},
the INITIAL program can be used to reconstruct a canonical
basis from the Picard--Fuchs equation of a single UT integral,
given sufficient derivative rank and an appropriate alphabet
ansatz~\cite{Dlapa:2020cwj}.
Geometric integration-by-parts orderings organize systems on maximal
cuts~\cite{e-collaboration:2025frv,Bree:2025tug}, while
Schwinger-parameter hypergeometric methods provide another route to
canonical bases~\cite{Jimenez-Santacruz:2026ecm}.

The above methods are mainly useful for polylog problems. Beyond that, for non-polylog problems, there are also various methods. Canonical constructions use period normalization and
algorithmic basis transformations~\cite{Adams:2018yfj,Frellesvig:2021hkr,Gorges:2023zgv}.
The extended INITIAL program can be used to construct canonical
differential equations for elliptic Feynman integrals~\cite{Dlapa:2022wdu}.
Other approaches use generalized integrand methods~\cite{Chen:2025hzq,Chaubey:2025adn,Yang:2025ofz,Forner:2026vby}
and Calabi--Yau geometry~\cite{Pogel:2022vat,Duhr:2025lbz,Maggio:2025jel}.
Their kernels can extend beyond algebraic $d\log$ forms, and the
integrals can have nonuniform weight~\cite{Frellesvig:2023iwr}.
Although we have the above non-polylog methods, in this paper, we focus on the logarithmic canonical form.

In practice, the above methods have different inputs, applicability
conditions and computational tools. One must choose a suitable combination of the tools for a given problem. After applying the tools, one uses algebraic diagnostics to see whether the tools worked to generate a valid CDE with $\varepsilon$ factorized. If not, for example, when there is a rank deficiency or failure of regulator factorization occurring in the DE, one has to revise the candidate integrals in the basis or the transformation ansatzes. This workflow is search-heavy, requiring many rounds of repeated attempts and expertise. But it is verification-light, checking whether there is a good CDE. Thus, this problem is well suited to AI agent-assisted automation.

In this paper, we present CANON, an AI agent-assisted framework that coordinates
UT basis constructions. The framework is based on a sector-by-sector approach. It retains the cumulative basis
transformation relative to the ordered source masters, allowing
integral proposals and direct differential-equation methods to be
combined. Guided by method cards, sector geometry, and algebraic
diagnostics, the agent selects methods and proposes candidate integral basis for the particular sector. 
Then the agent launches symbolic programs to carry out the calculations. After that, it uses independent verification tools to check whether the resulting transformation is valid. 

Once verified, the transformations will be stored together with their corresponding data, which supports the calculation of later sectors and subsequent sessions. After finishing all sector-independent searches, the agent combines the total transformations and verify the final results, containing canonical basis candidates and the corresponding CDE.

This paper is organized as follows. In Section \ref{sec:review}, we briefly introduce the concepts including Feynman integrals, the canonical basis, and CDE. In Section \ref{sec:agent-workflow}, we go through the CANON workflow and some of its important detailed design. In Section \ref{sec:applications} we take several examples to demonstrate how CANON workflow find and verify canonical basis. In Section \ref{sec:discussion} we provide some discussion and outlooks.

\Needspace{8\baselineskip}
\section{Review of Concepts and Methods}
\label{sec:review}

\subsection{From integral families to differential equations}
\label{sec:family-de}
 
Consider an \(L\)-loop integral family with loop momenta \(k_\ell\),
external momenta \(p\), and independent kinematic variables
\(\bm x=(x_1,\ldots,x_m)\). In dimensional regularization,
\(d=d_0-2\varepsilon\), with fixed reference dimension \(d_0\). An index
notation for its members is
\begin{equation}
  I_{\bm a}(\bm x,\varepsilon)
  =\int\prod_{\ell=1}^{L}\frac{\mathrm d^d k_\ell}{i\pi^{d/2}}
    \prod_{j=1}^{N}D_j^{-a_j}.
  \label{mc:family}
\end{equation}
The \(D_j\) are inverse propagators or auxiliary scalar products.
The integer indices \(a_j\) specify their powers, with nonpositive
powers encoding numerators. The propagators carry the causal prescription,
and overall normalization is part of the family definition.
The kinematic relations fix their dependence on \(\bm x\).
Integrals with the same set of propagators appearing
with positive powers form a sector. A subsector uses a (proper)
subset of those propagators.

IBP identities relate members of the
family~\cite{Chetyrkin:1981qh}. Using the identities, we can use reduction processes to express the required integrals
in a selected ordered master basis \(\bm I\). Differentiating the
masters and reducing the resulting integrals gives the source
differential equations (DEs)~\cite{Gehrmann:1999as}
\begin{equation}
  \partial_a\bm I=A_a\bm I,\qquad
  \partial_a=\frac{\partial}{\partial x_a}.
  \label{mc:source-de}
\end{equation}
The source system is specified by the family definition, ordered
masters, reduction data and derivatives in all requested kinematic
directions.

\subsection{Canonical form and basis conventions}
\label{sec:canonical-target}
\label{sec:scope}
 
One of the key tasks in the CDE method is to find a basis \(\bm J\) whose differential equations
are in canonical forms. It is equivalent to finding a basis transformation from a certain basis \(\bm I\) which is originally not a canonical basis. The transformation can be described by a
generically invertible matrix \(T(\bm x,\varepsilon)\), with
\(\det T\not\equiv0\), such that
\begin{equation}
  \bm I=T\bm J,\qquad
  B_a=T^{-1}(A_aT-\partial_aT).
  \label{mc:gauge}
\end{equation}
The transformed system is the \(\partial_a\bm J=B_a\bm J\). Notice that the functions allowed in a basis
change is specified by the coefficient domain specifies the, such as rational functions of \((\bm x,\varepsilon)\) with
algebraic extensions when needed. Also, the transformation may have poles or a vanishing determinant at
\(\varepsilon=0\) while remaining generically invertible.

Regulator factorization means that \(B_a/\varepsilon\) is independent
of \(\varepsilon\). The canonical target considered here further
requires a common logarithmic form~\cite{Henn:2013pwa}
\begin{equation}
  B_a=\varepsilon\sum_k R_k\,\partial_a\log W_k(\bm x),
  \label{mc:dlog}
\end{equation}
with regulator-independent letters \(W_k\) and matrices \(R_k\)
independent of both kinematics and regulator. The letters form the
logarithmic alphabet.

Eq.~\eqref{mc:dlog} determines the kinematic dependence order by
order in \(\varepsilon\). A basis of integrals with uniform transcendental weight (UT) \cite{Henn:2013pwa},
\begin{equation}
J_i=\varepsilon^k \sum_{n=0}^{\infty} \varepsilon^n J_i^{(n)}\,,
\end{equation}
where $J_i^{(n)}$ and is pure and has the transcendental weight $n$ \cite{Henn:2013pwa}, satisfies a canonical differential equation.

\subsection{Sector hierarchy and construction strategy}
\label{sec:proposal-verification}

When analyzing or transforming the DEs, the sector-by-sector approach is often used. If the master integrals are ordered by their sectors, with subsectors first, the source connection matrices are of block-lower-triangular forms~\cite{Meyer:2016slj}. In this paper we denote \(\overline{\bm I}_c\) as the \emph{sector cut} master integrals in the current sector, treating the sub-sector integrals as zero. The diagonal sector system of the cut master integrals, referred to below as the
\emph{core} of this sector in this paper, is then
\begin{equation}
  \partial_a\overline{\bm I}_c=A_{c,a}\overline{\bm I}_c,
  \label{mc:cut-de}
\end{equation}
A sector's core transformation \(T_c\) is defined as
\(\overline{\bm I}_c=T_c\overline{\bm K}_c\). The goal of a core transformation is to make
\(\partial_a\overline{\bm K}_c=\varepsilon F_{c,a}\overline{\bm K}_c\),
with \(F_{c,a}\) independent of \(\varepsilon\).

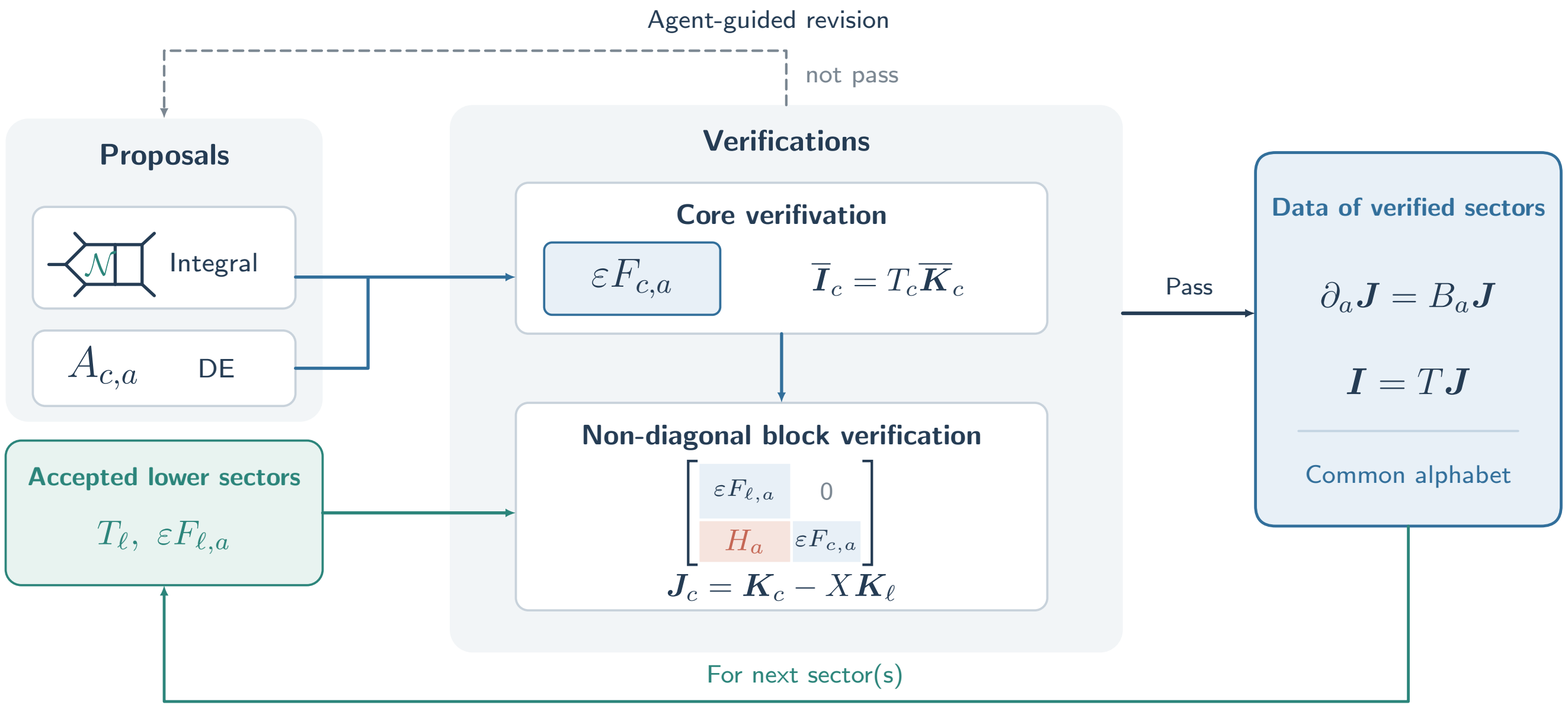
\begin{figure*}[t]
  \centering
  \resizebox{\textwidth}{!}{
\begingroup
\definecolor{CanonNavy}{HTML}{253C54}%
\definecolor{CanonBlue}{HTML}{326F9B}%
\definecolor{CanonTeal}{HTML}{2C857B}%
\definecolor{CanonRed}{HTML}{C56755}%
\definecolor{CanonGray}{HTML}{7B8792}%
\definecolor{CanonEdge}{HTML}{C8D2DB}%
\definecolor{CanonLight}{HTML}{F2F5F7}%
\definecolor{CanonPBlue}{HTML}{E8F0F7}%
\definecolor{CanonPTeal}{HTML}{E8F3F0}%
\definecolor{CanonPRed}{HTML}{F6E4DE}%
\begin{tikzpicture}[x=1bp,y=1bp,inner sep=0pt,outer sep=0pt,
 every node/.style={text=CanonNavy,font=\sffamily\fontsize{8.5}{10.2}\selectfont},
 panel/.style={draw=none,fill=CanonLight,rounded corners=7.2bp},
 card/.style={draw=CanonEdge,fill=white,line width=.7bp,rounded corners=4bp},
 title/.style={font=\sffamily\bfseries\fontsize{9}{10.8}\selectfont},
 flow/.style={-{Latex[length=3.3bp,width=2.8bp]},line width=1bp,line cap=round,line join=round},
 feedback/.style={flow,draw=CanonGray,dash pattern=on 3bp off 2bp,line width=.8bp}]
\path[use as bounding box] (0,-16) rectangle (507.6,214.56);
\begin{pgfinterruptboundingbox}
\path[panel] (5.76,80) rectangle (106.56,176.4);
\coordinate (proposals-north) at (56.16,176.4);
\node[title,font=\sffamily\bfseries\fontsize{9.5}{11.4}\selectfont]
 at (56.16,164.16) {Proposals};
\path[card] (14.4,116) rectangle (97.92,148.32);
\path[card] (14.4,85) rectangle (97.92,109);
\coordinate (integral-east) at (97.92,126);
\coordinate (de-east) at (97.92,97);
\node at (72,130) {Integral};
\node[font=\sffamily\fontsize{14}{16.8}\selectfont] at (36.72,97) {$A_{c,a}$};
\node at (72.72,97) {DE};
\begin{scope}[shift={(0,3.28)}]
\path[draw={rgb,1:red,0.1451;green,0.2353;blue,0.3294},fill=none,line width=1bp,line join=round,line cap=round] (31.158,133.038) -- (40.5864,133.038);
\path[draw={rgb,1:red,0.1451;green,0.2353;blue,0.3294},fill=none,line width=1bp,line join=round,line cap=round] (24.84,126.72) -- (31.158,133.038);
\path[draw={rgb,1:red,0.1451;green,0.2353;blue,0.3294},fill=none,line width=1bp,line join=round,line cap=round] (31.158,120.402) -- (24.84,126.72);
\path[draw={rgb,1:red,0.1451;green,0.2353;blue,0.3294},fill=none,line width=1bp,line join=round,line cap=round] (40.5864,120.402) -- (31.158,120.402);
\path[draw={rgb,1:red,0.1451;green,0.2353;blue,0.3294},fill=none,line width=1bp,line join=round,line cap=round] (49.14,120.402) -- (40.5864,120.402);
\path[draw={rgb,1:red,0.1451;green,0.2353;blue,0.3294},fill=none,line width=1bp,line join=round,line cap=round] (49.14,133.038) -- (49.14,120.402);
\path[draw={rgb,1:red,0.1451;green,0.2353;blue,0.3294},fill=none,line width=1bp,line join=round,line cap=round] (40.5864,133.038) -- (49.14,133.038);
\path[draw={rgb,1:red,0.1451;green,0.2353;blue,0.3294},fill=none,line width=1bp,line join=round,line cap=round] (40.5864,133.038) -- (40.5864,120.402);
\path[draw={rgb,1:red,0.1451;green,0.2353;blue,0.3294},fill=none,line width=1bp,line join=round,line cap=round] (31.158,133.038) -- (27.756,136.6344);
\path[draw={rgb,1:red,0.1451;green,0.2353;blue,0.3294},fill=none,line width=1bp,line join=round,line cap=round] (24.84,126.72) -- (19.494,126.72);
\path[draw={rgb,1:red,0.1451;green,0.2353;blue,0.3294},fill=none,line width=1bp,line join=round,line cap=round] (31.158,120.402) -- (27.756,116.8056);
\path[draw={rgb,1:red,0.1451;green,0.2353;blue,0.3294},fill=none,line width=1bp,line join=round,line cap=round] (49.14,120.402) -- (53.028,116.8056);
\path[draw={rgb,1:red,0.1451;green,0.2353;blue,0.3294},fill=none,line width=1bp,line join=round,line cap=round] (49.14,133.038) -- (53.028,136.6344);
\node[anchor=center,text={rgb,1:red,0.1725;green,0.5216;blue,0.4824},font={\sffamily\fontsize{10}{12}\selectfont}] at (36.018,126.72) {$\mathcal{N}$};
\end{scope}
\path[draw=CanonTeal,fill=CanonPTeal,line width=.7bp,rounded corners=5bp]
 (5.76,28) rectangle (106.56,74);
\coordinate (lower-east) at (106.56,51);
\coordinate (lower-south) at (56.16,28);
\node[text=CanonTeal,font=\sffamily\bfseries\fontsize{8.3}{9.96}\selectfont]
 at (56.16,61.8) {Accepted lower sectors};
\node[text=CanonTeal,font=\sffamily\fontsize{11}{13.2}\selectfont]
 at (56.16,43.2) {$T_\ell,\ \varepsilon F_{\ell,a}$};
\path[panel] (147,6.6) rectangle (361,180.72);
\coordinate (check-north) at (254,180.72);
\node[title,font=\sffamily\bfseries\fontsize{9.5}{11.4}\selectfont]
 at (254,169.5) {Verifications};
\path[card] (168,108) rectangle (337,156);
\coordinate (core-west) at (168,126);
\coordinate (core-south) at (252.5,108);
\node[title] at (252.5,146) {Core verifivation};
\path[draw=CanonBlue,fill=CanonPBlue,line width=.7bp,rounded corners=2.52bp]
 (177,114) rectangle (233,137);
\node[font=\sffamily\fontsize{13}{15.6}\selectfont] at (205,125.5) {$\varepsilon F_{c,a}$};
\node[font=\sffamily\fontsize{10.3}{12.36}\selectfont] at (286.5,125.5)
 {$\overline{\boldsymbol{I}}_c=T_c\overline{\boldsymbol{K}}_c$};
\begin{scope}[yshift=-8bp]
\path[card] (168,28) rectangle (337,94);
\coordinate (combined-north) at (252.5,94);
\coordinate (combined-west) at (168,59);
\coordinate (combined-east) at (337,59);
\node[title] at (252.5,83) {Non-diagonal block verification};
\path[fill=CanonPBlue,draw=white,line width=.8bp] (226,56.76) rectangle (255.64,75);
\path[fill=CanonPBlue,draw=white,line width=.8bp] (255.64,43) rectangle (278,56.76);
\path[fill=CanonPRed,draw=white,line width=.8bp] (226,43) rectangle (255.64,56.76);
\draw[CanonNavy,line width=.8bp] (225.8,75.5) -- (222.8,75.5) -- (222.8,42.5) -- (225.8,42.5);
\draw[CanonNavy,line width=.8bp] (278.2,75.5) -- (281.2,75.5) -- (281.2,42.5) -- (278.2,42.5);
\node[font=\sffamily\fontsize{7.7}{9.24}\selectfont] at (240.82,65.88) {$\varepsilon F_{\ell,a}$};
\node[font=\sffamily\fontsize{7.7}{9.24}\selectfont] at (266.82,49.88) {$\varepsilon F_{c,a}$};
\node[text=CanonRed,font=\sffamily\fontsize{9.6}{11.52}\selectfont] at (240.82,49.88) {$H_a$};
\node[text=CanonGray,font=\sffamily\fontsize{8}{9.6}\selectfont] at (266.82,65.88) {0};
\node[font=\sffamily\fontsize{10.2}{12.24}\selectfont] at (252.5,35.3)
 {$\boldsymbol{J}_c=\boldsymbol{K}_c-X\boldsymbol{K}_\ell$};
\end{scope}
\draw[CanonBlue,line width=1bp,line cap=round,line join=round]
 (de-east) -- (121,97) -- (121,126);
\draw[flow,draw=CanonBlue] (integral-east) -- (core-west);
\draw[flow,draw=CanonBlue] (core-south) -- (combined-north);
\draw[flow,draw=CanonTeal] (lower-east) -- (combined-west);
\path[draw={rgb,1:red,0.1961;green,0.4353;blue,0.6078},fill={rgb,1:red,0.9098;green,0.9412;blue,0.9686},line width=0.9bp,line join=round,line cap=round] (409.68,46.8) -- (493.92,46.8) .. controls (498.24,46.8) and (500.4,48.96) .. (500.4,53.28) -- (500.4,159.12) .. controls (500.4,163.44) and (498.24,165.6) .. (493.92,165.6) -- (409.68,165.6) .. controls (405.36,165.6) and (403.2,163.44) .. (403.2,159.12) -- (403.2,53.28) .. controls (403.2,48.96) and (405.36,46.8) .. (409.68,46.8) -- cycle;
\path[draw={rgb,1:red,0.7647;green,0.8314;blue,0.8863},fill=none,line width=0.7bp,line join=round,line cap=round] (416.88,77.04) -- (486.72,77.04);
\node[anchor=center,text={rgb,1:red,0.1961;green,0.4353;blue,0.6078},font={\sffamily\bfseries\fontsize{8.2}{11.04}\selectfont}] at (451.8,148.32) {Data of verified sectors};
\node[anchor=center,text={rgb,1:red,0.1451;green,0.2353;blue,0.3294},font={\sffamily\fontsize{11.2}{13.44}\selectfont}] at (451.8,119.52) {$\partial_a\boldsymbol{J}=B_a\boldsymbol{J}$};
\node[anchor=center,text={rgb,1:red,0.1451;green,0.2353;blue,0.3294},font={\sffamily\fontsize{12}{14.4}\selectfont}] at (451.8,92.88) {$\boldsymbol{I}=T\boldsymbol{J}$};
\node[anchor=center,text={rgb,1:red,0.1961;green,0.4353;blue,0.6078},font={\sffamily\fontsize{8.1}{9.72}\selectfont}] at (451.8,62.64) {Common alphabet};
\coordinate (verified-west) at (403.2,114.48);
\coordinate (verified-south) at (451.8,46.8);
\coordinate (check-east) at (361,114.48);
\draw[flow,draw=CanonNavy,line width=1.15bp] (check-east) -- (verified-west);
\node[font=\sffamily\fontsize{7.7}{9.24}\selectfont] at (382.1,123.2) {Pass};
\draw[feedback] (check-north) -- (254,198) -- (56.16,198) -- (proposals-north);
\node[anchor=west,text=CanonGray,font=\sffamily\fontsize{8.1}{9.72}\selectfont]
 at (260,189.4) {not pass};
\node[font=\sffamily\fontsize{8.1}{9.72}\selectfont] at (248.4,207.36) {Agent-guided revision};
\draw[flow,draw=CanonTeal,line width=.9bp]
 (verified-south) -- (451.8,-9) -- (56.16,-9) -- (lower-south);
\node[text=CanonTeal,font=\sffamily\fontsize{8}{9.6}\selectfont] at (260,-1) {For next sector(s)};
\end{pgfinterruptboundingbox}
\end{tikzpicture}
\endgroup}
  \caption{The working procedure in a certain sector.
  Integral representatives and direct DE transformations feed the core
  check. The checked core is combined with the accepted lower-sector
  transformation to complete the coupled system. Blue blocks have
  epsilon form; the red coupling requires a lower-sector correction.
  Verification checks the cumulative transformation and a common
  logarithmic alphabet. Dashed feedback guides revision, while accepted
  results supply the lower-sector input for subsequent sectors.
  The integral and matrix blocks are schematic.}
  \label{fig:scientific-workflow}
\end{figure*}

\Needspace{6\baselineskip}
\section{The CANON Agent Workflow}
\label{sec:agent-workflow}

The CANON workflow runs over all integral sectors, from bottom to top, by default. When performing in a given sector, it first makes a proposal, containing the candidate integrals in the current sector and the corresponding DE. Then, it verifies them. The verification includes two aspects. Firstly, it checks the core system corresponding to the sector cut integrals, requiring it to be $\varepsilon$ factorized and fully ranked. Secondly, it checks the couplings of the candidate integrals with sub-sector integrals, requiring the non-diagonal blocks also being $\varepsilon$ factorized. If the verification steps pass, the current sector is considered finished. Otherwise, the agent revises the proposal and revise again the revised proposal. If a sector is finished, necessary information during its procedure is stored to support the processes for later super sectors. Once all requested sectors are finished, CANON runs a final check to test the transformation validity and \rm{d-log} property of the CDE and determine whether the project is accomplished. Figure~\ref{fig:scientific-workflow} summarizes this workflow.

\subsection{Coordination of main and worker agents}
\label{sec:orchestration}
 
CANON provides a main agent interacting with its user, coordinate the calculation and selects the next step
from the current results, with the support of status arrangement program. It assigns each worker agent a specific problem,
input files and conventions, and a resource budget. Workers apply
selected methods and return their proposal on candidate basis. The main agent then tests the candidates and decides
whether the results are ready for use in subsequent calculations.

\Needspace{5\baselineskip}
\subsection{Method cards in the agent workflow}
\label{sec:routing}

A method card is a scientific reference describing (usually a SKILL and some related scripts for the agent) a certain basis construction methods,
its input conventions, applicability conditions and implemented scripts. Within an assignment, worker agents use this information with
the current diagnostics to choose or revise a basis construction attempt.
Several cards can address the same unresolved problem, and worker agents will determine which one to use by themselves, under some human-guided instructions.
Proposals from different methods can be
composed and checked against the same source system. These method prompts are independent of the agent model and host. However, changing underlying language model will vary the performance of our workflow when proposing integral candidates, as different models have their own tendency when picking method cards, and therefore make a difference to the final results. Generally, we suggest our user to apply stronger language models in CANON workflow to guarantee the basis could be found efficiently.

\Needspace{5\baselineskip}
\subsection{Persistent research state and artifact reuse}
\label{sec:persistence}
Long-term researches like canonical basis construction will easily exhaust the context of AI agent, and cause workflow performance drop. Thus, sub-agents and persistent state management are needed for these jobs. Our persistent record comprises a directed acyclic task graph,
execution attempts and registered artifacts, so that can provide an external long-term arrangement for CANON workflow. In the task graph, each node specifies a calculation task and the
criteria for accepting the result. A task
becomes ready when all direct prerequisites have been accepted, written
\(\mathcal D_i\subseteq\mathcal C\), where \(\mathcal D_i\) is its prerequisite set and \(\mathcal C\) is
the set of completed tasks. The coordinator extends this graph as the
calculation develops.

\begin{figure*}[t]
  \centering
  \resizebox{\textwidth}{!}{\input{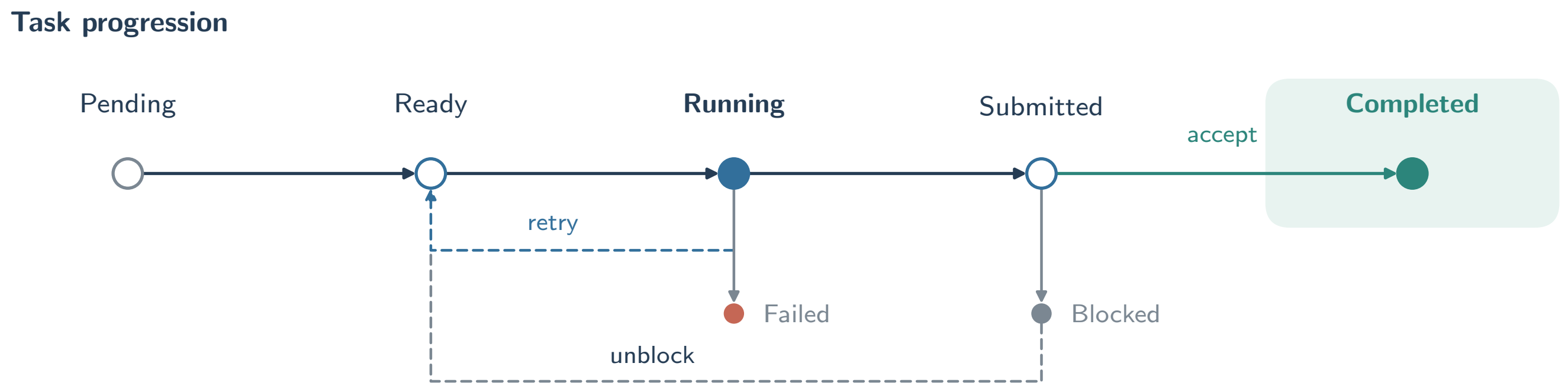}}
  \caption{
    Persistent research state and artifact reuse in CANON. This figure shows the task progression of CANON workflow. 
  }
  \label{fig:persistent-research-state}
\end{figure*}

Figure~\ref{fig:persistent-research-state} shows the progression from
readiness to completion. Each worker claim creates a separate execution
record and working directory. The main agent reviews the submitted calculation and its verification
results, then accepts it if they meet the task criteria. Acceptance
makes the result available to dependent tasks. The state manager
records this decision and updates their readiness.

Failed attempts will be rejected by the main agent and retain their diagnostics and artifacts. A retry returns
the task to ready; otherwise it remains
failed. An external condition (namely the computation resources exhausted or human instruction) can block a task until the main agent
confirms that work can resume. Previously verified lower-sector bases remain available during
retries and interruptions.

Artifacts are stored by reference with content hashes, including ordered
masters, differential equations, reduction rules and cumulative
transformations. If \(\mathcal A_j^\star\) denotes the artifacts of
the accepted execution of task \(j\), task \(i\) reuses
\(\mathcal I_i=\bigcup_{j\in\mathcal D_i}\mathcal A_j^\star\).
New or modified results are registered separately. 

\Needspace{5\baselineskip}
\subsection{Modular design and extensions}
\label{sec:modularity}
To enhance the flexibility and efficiency on solving the most cutting-edge problems, CANON separates most of its working procedures, including kinematic parameterization, master discovery, IBP reduction, basis
construction and verification into pluggable components.
Each component supplies scientific instructions and communicates through
shared settings that preserve basis order, normalization and the
cumulative transformation. Components can be replaced or combined within the same agent workflow, and independent routines check their outputs.

For some complicated problems, additional powerful plugins may play important roles. In Section \ref{sec:big-example}, we present an example about this. 

\subsection{The outputs of CANON workflow}

The CANON workflow can provide results in the format defined by the user. By default, CANON will provide a collection of canonical basis (in Mathematica expression format) and a PDF report that contains a summary of workflow along with a detailed list, demonstrating how CANON find each canonical basis candidates sector by sector. In the PDF report, the citations and corresponding links are included for the methods CANON used, making it convenient for users to cite them in their research works. Besides the default outputs, users can also let CANON to perform analytical reconstruction or use other user-defined packages to obtain a full analytic CDE matrix and the corresponding alphabet.

\section{Examples}
\label{sec:applications}

We demonstrate some examples in this section. For the examples in Section \ref{sec:offshell-doublebox}, \ref{sec:pentabox} and \ref{sec:three-loop}, all canonical basis and corresponding CDEs are independently constructed by public CANON workflow. Table~\ref{tab:case-systems}
summarizes these examples and their verification. For the example in \ref{sec:big-example}, we introduced additional private modules into the workflow to obtain the full family results.

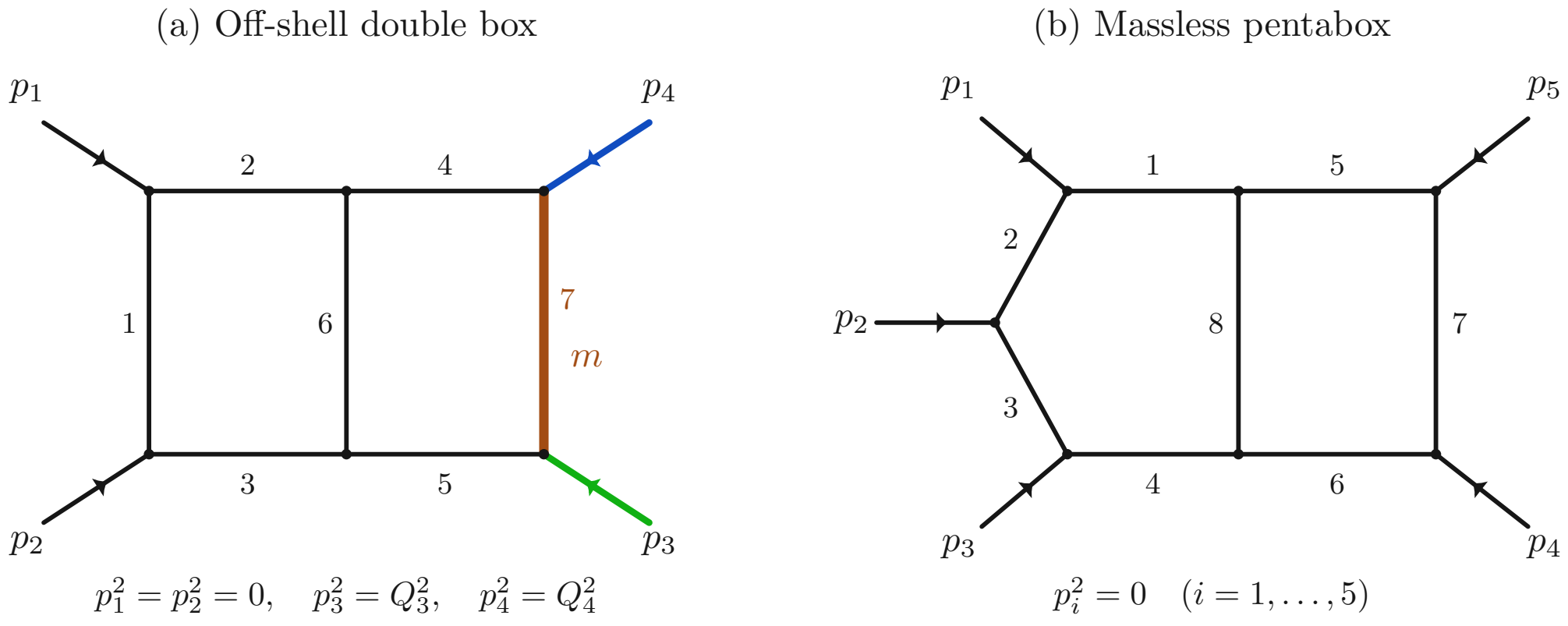
\begin{figure*}[t]
  \centering
  \resizebox{0.8\textwidth}{!}{
\begingroup
\definecolor{TopologyInk}{RGB}{22,22,22}
\definecolor{TopologyMass}{RGB}{162,76,19}
\definecolor{ColorQ3}{RGB}{16,176,19}
\definecolor{ColorQ4}{RGB}{16,76,192}
\begin{tikzpicture}[x=1bp,y=1bp,inner sep=0pt,outer sep=0pt,
 every node/.style={text=TopologyInk,font=\rmfamily\fontsize{11}{13}\selectfont},
 propagator/.style={draw=TopologyInk,line width=1.35bp,line cap=round},
 massive/.style={draw=TopologyMass,line width=2.8bp,line cap=round},
 momentum/.style={propagator,postaction={decorate},decoration={markings,
 mark=at position .60 with {\arrow{Latex[length=3.2bp,width=6.8bp]}}}},
 number/.style={font=\rmfamily\fontsize{9}{11}\selectfont}]
\path[use as bounding box] (0,0) rectangle (487.97049,198.846);
\begin{pgfinterruptboundingbox}
\coordinate (db-tl) at (50.4315,134.6969);
\coordinate (db-bl) at (50.4315,56.3105);
\coordinate (db-tm) at (109.2213,134.6969);
\coordinate (db-bm) at (109.2213,56.3105);
\coordinate (db-tr) at (168.0111,134.6969);
\coordinate (db-br) at (168.0111,56.3105);
\draw[propagator] (db-tl) -- (db-bl);
\draw[propagator] (db-tl) -- (db-tm);
\draw[propagator] (db-bl) -- (db-bm);
\draw[propagator] (db-tm) -- (db-tr);
\draw[propagator] (db-bm) -- (db-br);
\draw[propagator] (db-tm) -- (db-bm);
\draw[massive] (db-tr) -- (db-br);
\foreach \x/\y/\t in {44.547/95.395/1,79.821/142.427/2,79.821/47.580/3,138.611/142.427/4,138.611/47.580/5,102.945/95.395/6}
 \node[number] at (\x,\y) {\t};
\node[number,text=TopologyMass] at (175.061,102.450) {7};
\node[text=TopologyMass] at (180.774,84.766) {$m$};
\draw[momentum] (19.0769,155.0773) -- (db-tl);
\draw[momentum] (19.0769,35.9300) -- (db-bl);
\draw[momentum,line width=2pt,color=ColorQ3] (199.3657,35.9300) -- (db-br);
\draw[momentum,line width=2pt,color=ColorQ4] (199.3657,155.0773) -- (db-tr);
\node at (14.3,164.4) {$p_1$};
\node at (14.3,29.6) {$p_2$};
\node at (202.5,29.6) {$p_3$};
\node at (202.5,164.4) {$p_4$};
\foreach \v in {db-tl,db-bl,db-tm,db-bm,db-tr,db-br}
 \fill[TopologyInk] (\v) circle[radius=1.55bp];
\node at (109.2213,183.6) {(a) Off-shell double box};
\node[font=\rmfamily\fontsize{10}{12}\selectfont] at (109.2213,14.3)
 {$p_1^2=p_2^2=0,\quad p_3^2=Q_3^2,\quad p_4^2=Q_4^2$};
\coordinate (pb-tl) at (323.8787,134.6969);
\coordinate (pb-lm) at (302.3225,95.5037);
\coordinate (pb-bl) at (323.8787,56.3105);
\coordinate (pb-tm) at (374.8299,134.6969);
\coordinate (pb-bm) at (374.8299,56.3105);
\coordinate (pb-tr) at (433.6197,134.6969);
\coordinate (pb-br) at (433.6197,56.3105);
\draw[propagator] (pb-tl) -- (pb-tm);
\draw[propagator] (pb-tl) -- (pb-lm);
\draw[propagator] (pb-lm) -- (pb-bl);
\draw[propagator] (pb-bl) -- (pb-bm);
\draw[propagator] (pb-tm) -- (pb-tr);
\draw[propagator] (pb-bm) -- (pb-br);
\draw[propagator] (pb-tr) -- (pb-br);
\draw[propagator] (pb-tm) -- (pb-bm);
\foreach \x/\y/\t in {349.349/142.427/1,307.020/120.479/2,307.020/70.312/3,349.349/47.580/4,404.220/142.427/5,404.220/47.580/6,440.669/95.395/7,368.162/95.395/8}
 \node[number] at (\x,\y) {\t};
\draw[momentum] (298.4031,156.2531) -- (pb-tl);
\draw[momentum] (267.0486,95.5037) -- (pb-lm);
\draw[momentum] (298.4031,34.7542) -- (pb-bl);
\draw[momentum] (461.0549,34.7542) -- (pb-br);
\draw[momentum] (461.0549,156.2531) -- (pb-tr);
\node at (291.7,165.2) {$p_1$};
\node[anchor=east] at (265,95.5) {$p_2$};
\node at (291.7,28.8) {$p_3$};
\node at (466.1,28.8) {$p_4$};
\node at (466.1,165.2) {$p_5$};
\foreach \v in {pb-tl,pb-lm,pb-bl,pb-tm,pb-bm,pb-tr,pb-br}
 \fill[TopologyInk] (\v) circle[radius=1.55bp];
\node at (367.0,183.6) {(b) Massless pentabox};
\node[font=\rmfamily\fontsize{10}{12}\selectfont] at (367.0,14.3)
 {$p_i^2=0\quad(i=1,\ldots,5)$};
\end{pgfinterruptboundingbox}
\end{tikzpicture}
\endgroup}
  \caption{Top sectors of the two-loop examples. All external momenta are
  incoming. Internal labels give the propagator positions in the
  family definitions; auxiliary numerator denominators are omitted.
  In (a), line 7 has mass \(m\), and all other internal lines are
  massless. The external virtualities \(Q_3^2\) and \(Q_4^2\) are
  independent. In (b), all internal and external lines are massless.}
  \label{fig:example-topologies}
\end{figure*}

\subsection{An off-shell double box with one internal mass}
\label{sec:offshell-doublebox}

The first example is the planar double box family in
Fig.~\ref{fig:example-topologies}(a). The external and loop momenta are with dimension $d=4-2\varepsilon$. We use all-incoming momenta,
\(p_1+p_2+p_3+p_4=0\), with

\(p_1^2=p_2^2=0\), \(p_3^2=Q_3^2\) and \(p_4^2=Q_4^2\).
The independent variables are
\[
  \begin{gathered}
  \bm x=(s,t,Q_3^2,Q_4^2,m^2),\\
  s=(p_1+p_2)^2,\qquad t=(p_2+p_3)^2.
  \end{gathered}
\]
All five variables remain generic. The family contains 37 master
integrals in 24 nonvanishing sectors, including three masters in the
seven-propagator top sector. With loop momenta \(k_1,k_2\),
the physical inverse propagators are
\begin{equation}
\begin{aligned}
 D_1&=k_1^2,& D_2&=(k_1+p_1)^2,\\
 D_3&=(k_1-p_2)^2,& D_4&=(k_2+p_1)^2,\\
 D_5&=(k_2-p_2)^2,& D_6&=(k_1-k_2)^2,\\
 D_7&=(k_2-p_2-p_3)^2-m^2.
\end{aligned}
\label{eq:doublebox-propagators}
\end{equation}
The irreducible scalar-product (ISP) coordinates are
\begin{equation}
 D_8=k_2^2,\qquad D_9=(k_1-p_2-p_3)^2.
\end{equation}
Their indices obey \(a_8,a_9\leq0\), so they supply numerator
insertions. 

The completed connection uses 21 active logarithmic letters. Its
algebraic part contains the single square root
\[
 r=\sqrt{(s-Q_3^2-Q_4^2)^2-4Q_3^2Q_4^2}.
\]
Seventeen letters are rational in the five variables, and four have
the form \((a+r)/(a-r)\) where $a$ stands for certain polynomial of kinematic variables, listed below:
\begin{align*}
 a\in\{&s+Q_3^2-Q_4^2,\ s-Q_3^2+Q_4^2,\\
       &s-Q_3^2-Q_4^2+2m^2,\ s-Q_3^2-Q_4^2+2t\}.
\end{align*}

The cumulative transformation satisfies the exact gauge identity in all
five directions. The final verification also checks integrability conditions and reconstructs every differential matrix. 

The DE system of this example before and after applying CANON workflow is shown in the left column in Figure \ref{fig:example-de-maps}.

\subsection{The massless two-loop five-point pentabox}
\label{sec:pentabox}
 
The second example is the pentabox in Fig.~\ref{fig:example-topologies}(b) (where external and loop momenta are with dimension $d=4-2\varepsilon$). It is a standard
multiscale test with 61 master integrals in 46 sectors. Its canonical
system and pentagon functions are known~\cite{Gehrmann:2018yef}.
We keep the five adjacent invariants
\(\bm x=(s_{12},s_{23},s_{34},s_{45},s_{51})\), where
\(s_{ij}=(p_i+p_j)^2\), \(p_i^2=0\), and \(\sum_i p_i=0\).
The source system includes derivatives with respect to all five
invariants, including the overall scale direction.
With loop momenta \(k_1,k_2\), the eight physical inverse propagators
are
\begin{equation}
\begin{aligned}
 D_1&=k_1^2,&D_2&=(k_1+p_1)^2,\\
 D_3&=(k_1+p_1+p_2)^2,\\
 D_4&=(k_1+p_1+p_2+p_3)^2,\\
 D_5&=k_2^2,&D_6&=(k_2+p_1+p_2+p_3)^2,\\
 D_7&=(k_2-p_5)^2,&D_8&=(k_1-k_2)^2.
\end{aligned}
\label{eq:pentabox-propagators}
\end{equation}
The three ISP coordinates are
\begin{equation}
\begin{aligned}
 D_9&=(k_1-p_5)^2,&D_{10}&=(k_2+p_1)^2,\\
 D_{11}&=(k_2+p_1+p_2)^2,
\end{aligned}
\end{equation}
with \(a_9,a_{10},a_{11}\leq0\). The overall scale is restored
in the source equations even when the reduction is performed at
\(s_{51}=1\).

For this family, the final CDEs have 24 active letters. Reconstruction of all 46 sectors yields a 61-master basis. Its final
verification confirms the exact gauge identity, epsilon
factorization and integrability, and reconstructs all five
directional matrices. 

The DE system of this example before and after applying CANON workflow is shown in the middle column in Figure \ref{fig:example-de-maps}.

\subsection{A three-loop family with forty master integrals}
\label{sec:three-loop}
 
The third example is a three-loop example contains 40 master integrals in 28 sectors. The external and loop momenta are with dimension $d=4-2\varepsilon$
Let \(p^2=\ell^2=0\), \((p+\ell)^2=s\), and let \(u\) be dimensionless.
For loop momenta \(k_1,k_2,k_3\), the physical inverse propagators are
\begin{equation}
\begin{aligned}
 D_1&=k_1^2,&D_4&=\bigl(k_2+(1-u)p\bigr)^2,\\
 D_5&=(k_1+k_2)^2,&D_6&=(\ell+k_1+k_2)^2,\\
 D_8&=k_3^2,&D_9&=(\ell+k_2+k_3-up)^2,\\
 D_{11}&=(\ell-p+k_3)^2,\\
 D_{12}&=(k_1-k_3+up)^2.
\end{aligned}
\label{eq:three-loop-propagators}
\end{equation}

We keep the original indexing of the twelve quantities. The remaining
four are the ISP coordinates,
\begin{equation}
\begin{aligned}
 D_2&=(p+k_3)^2,&D_3&=(k_1+up)^2,\\
 D_7&=(k_2+k_3)^2,&D_{10}&=k_2^2,
\end{aligned}
\end{equation}
with \(a_2,a_3,a_7,a_{10}\leq0\). Thus the differential system has
one kinematic scale \(s\) and one dimensionless variable \(u\).
The canonical differential equation is
\begin{equation}
 \mathrm d\bm J=\varepsilon\bigl[
 -3\,\mathbf{1}_{40}\,\mathrm d\log s
 +R_u\,\mathrm d\log u
 +R_{1-u}\,\mathrm d\log(1-u)\bigr]\bm J,
 \label{eq:three-loop-example}
\end{equation}
where \(R_u\) and \(R_{1-u}\) are constant rational matrices. The active alphabet is only
\(\{s,u,1-u\}\) and the scale term separates from the ratio dependence.

The calculation used our DlogBasis-refine interface (see Appendix \ref{app:interfaces}) to build candidates to validate the core system, by varying loop order (in LBL Baikov representation), dots and irreducible numerators. We use CANON workflow to reconstruct the basis through
all 28 sectors. The resulting full 40-master system passed the validation check. 

\begin{figure*}[t]
 \centering
 \includegraphics[width=\textwidth]{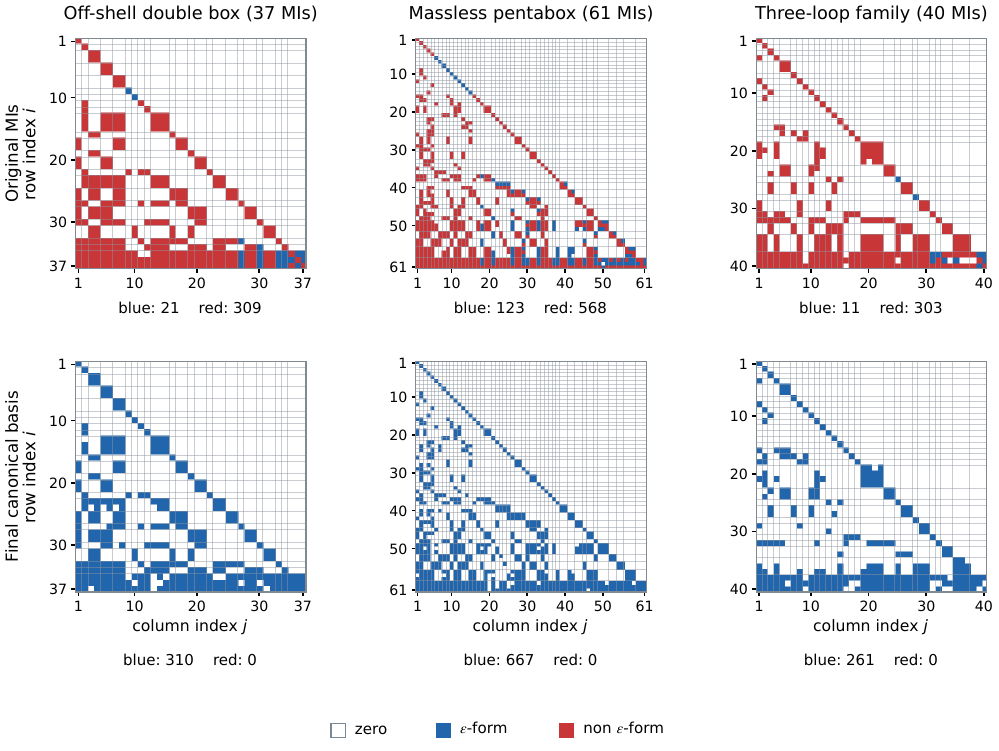}
 \caption{This figure compares the original master-integral
systems with their final canonical forms for examples in Section \ref{sec:offshell-doublebox}, \ref{sec:pentabox} and \ref{sec:three-loop}, respectively. Each square combines the
same matrix position over all differential directions.
White denotes a zero in every direction. A nonzero position is blue
if every directional entry divided by \(\varepsilon\) is independent
of \(\varepsilon\), and red otherwise. The absence of red entries in
the final panels displays epsilon factorization. The symbolic checks described above establish these identities at
generic kinematics.}
 \label{fig:example-de-maps}
\end{figure*}

The DE system of this example before and after applying CANON workflow is shown in the right column in Figure \ref{fig:example-de-maps}.

\begin{table*}[t]
\caption{Canonical systems for the three primary examples. All kinematic
variables and the regulator remain symbolic in the final checks.
The canonical connections use a shared set of constant residue matrices
in every differential direction.}
\label{tab:case-systems}
\begin{ruledtabular}
\begin{tabular}{lccc}
 & Off-shell double box & Massless pentabox & Three-loop family\\
\hline
Sectors & 24 & 46 & 28 \\
Master integrals & 37  & 61  & 40 \\
Variables & 5 & 5  & 2 \\
Letters in alphabet & 21 & 24 & 3\\
Verification & {PASS} & {PASS} & {PASS}
\end{tabular}
\end{ruledtabular}
\end{table*}

\subsection{An more complicated benchmark}
\label{sec:big-example}

\begin{figure*}[t]
 \centering
 \includegraphics[width=0.92\textwidth]{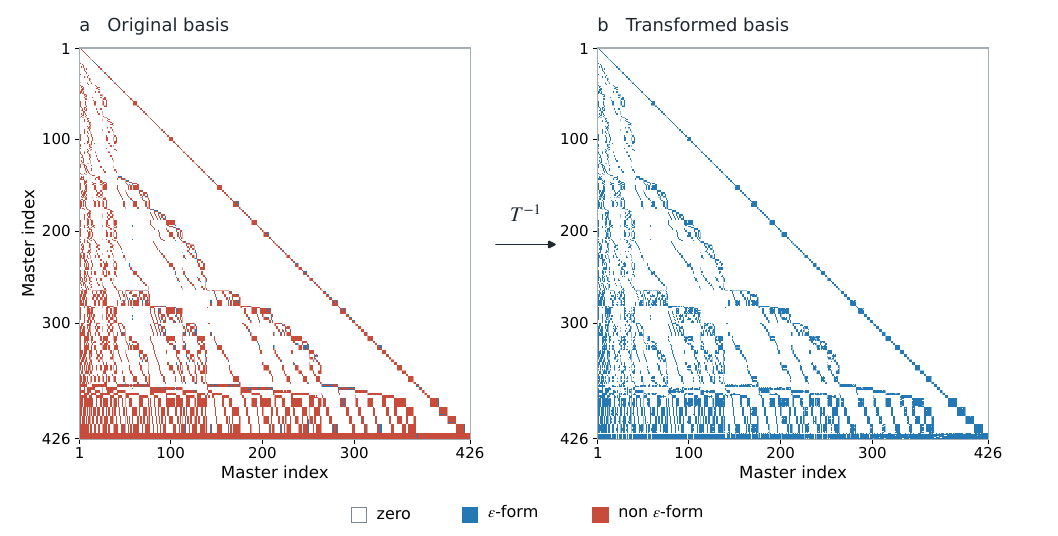}
 \caption{Matrix structure before and after the basis change
 \(\bm J=T^{-1}\bm I\) for a 426-master system.
 Both panels combine the same derivative directions at one fixed
 kinematic point, with the regulator kept symbolic.
 White denotes entries that vanish in every direction; blue denotes
 nonzero entries proportional to \(\varepsilon\) with
 \(\varepsilon\)-independent coefficients; red marks a failure of this
 condition in at least one direction. Both panels use the same sector ordering.}
 \label{fig:large-system-de}
\end{figure*}

In this subsection, we demonstrate a benchmark example. This family has 426 master integrals, which is much more than the previous three examples. All canonical basis are obtained and verified by CANON workflow with some additional private modules within 48 hours. Figure \ref{fig:large-system-de} shows the DE system before and after applying the CANON workflow.

This example is selected from an unpublished project. The researchers involved have approved the publication of the material presented here.

\section{Summary}
\label{sec:discussion}
In this article, we introduce CANON, an AI agent workflow which automatically proposes and verifies canonical basis and corresponding CDE. It combines various methods from professional experiences and well-developed packages to raise basis proposals, and check their validity using built-in verification modules in CANON. The completed transformations and their inputs are stored to support building later super sectors in the CANON workflow. The modular architecture of CANON workflow also enables users to plugin their self-developed methods or modules to tackle with more advanced canonical basis problems. 

The CANON results, by default, contain the canonical basis, and an report (in pdf format) summarizing the method the agent used together with the citations about the methods. The results can also contain the CDE and symbol letter alphabet if they are requested by the user.

One application of this modular approach would be to extend CANON to integral families with elliptic sectors. The two-loop kite family provides a concrete test, since it contains an elliptic sunrise subsector and its full \(\varepsilon\)-form is known~\cite{Adams:2018yfj}. Starting from a period-normalized basis for the sunrise, the agent could search for a compatible basis for the remaining integrals. We look forward to apply CANON workflow in future phenomenology researches and improve it spontaneously to enhance our understanding and performances on perturbative Feynman integral calculations.
\\[5pt]

\textbf{Declaration of AI usage}: 

Generative AI is used to assist in the drafting of this paper. All the content of this paper has gone through full human review and revision, for which the authors can take responsibility.

The development of CANON workflow is assisted with AI agents under human supervision. We kindly remind that, since the performance relies on the underlying models working in CANON, users should be aware of the possible performance change and potential risk due to the indetermination of the LLM outputs.

\section{Acknowledgment}
We thank Rourou Ma and Yongqun Xu for helpful discussions. Yuanche Liu is supported by NSFC through Grant No. 124B1014. Yu Wu is supported by ANR-24-CE31-799. Zihao Wu is supported by South China Normal University through grant No. 43102673.
Dong-Shan Jian, Yan-Qing Ma and Bo-Xuan Shi is in part supported by the National Natural Science Foundation of China (No. 12325503). Yang Zhang is supported by the National Natural Science Foundation of China (No. 12575078 and 12247103).

\begin{table*}[htbp]

  \caption{Some details on different method cards employed in CANON workflow}
  \label{mc:provider-table}
  \begin{ruledtabular}
  \begin{tabular}{p{0.19\textwidth}p{0.31\textwidth}p{0.41\textwidth}}
    Construction & Construction principle & Input and scope in CANON \\
    \hline
    {\raggedright Integral-based construction\par} &
    {\raggedright Geometry and singularities guide integral proposals and
    normalization. IBP reduction expresses the proposals in the source
    master basis, retaining lower-sector coefficients for
    completion~\cite{Henn:2020lye,Dlapa:2021qsl,Chetyrkin:1981qh,He:2022ctv}\par} &
    {\raggedright Defined integral family and kinematics, finite numerator
    and propagator-power bounds, normalization hypotheses, and reduction
    data for the proposed integrals\par} \\
    {\raggedright CANONICA\par} &
    {\raggedright Rational gauge construction by regulator expansion
    and recursive sector transformation~\cite{Meyer:2016slj,Meyer:2017joq}\par} &
    {\raggedright Rational source differential equations in one or more
    kinematic variables\par} \\
    {\raggedright \(\varepsilon=0\) polynomial construction\par} &
    {\raggedright Finite homogeneous-polynomial search for constant combinations of the
    \(\varepsilon=0\) system, with rational or algebraic normalization prefactors\par} &
    {\raggedright Rational source system regular at \(\varepsilon=0\),
    with any required algebraic generators supplied\par} \\
    {\raggedright Libra\par} &
    {\raggedright Fuchsification (reduction to simple poles), residue normalization
    and regulator factorization~\cite{Lee:2014ioa,Lee:2020zfb}\par} &
    {\raggedright Small rational systems in one variable. Multivariable
    input requires an explicit finite directional proposal ending in
    jointly Fuchsian form\par} \\
    {\raggedright Residue-balance construction\par} &
    {\raggedright Bounded rational shears, integer shifts of residue
    eigenvalues and regulator factorization~\cite{Lee:2014ioa}\par} &
    {\raggedright One nontrivial variable, possibly after exact scale
    separation, with rational pole locations independent of \(\varepsilon\)
    and residue eigenvalues \(n+c\varepsilon\), \(n\in\mathbb Z\)\par}
  \end{tabular}
  \end{ruledtabular}
\end{table*}

\section*{Data Availability}
The CANON source code and documented interfaces are maintained at
\url{https://github.com/Alice-Shimada/CANON-release}.
The accompanying archive contains the canonical bases, logarithmic
connections and final verification records for the three
examples in Section \ref{sec:offshell-doublebox}, \ref{sec:pentabox} and \ref{sec:three-loop}.

\newpage 
\appendix

\par\noindent
\section{Scientific Interfaces and Execution Details}
\label{app:interfaces}

The default reduction interfaces in CANON workflow is
Kira3~\cite{Maierhofer:2017gsa,Klappert:2020nbg,Lange:2025fba}. FIRE7~\cite{Smirnov:2025prc}, NeatIBP ~\cite{Wu:2023upw,Wu:2025aeg}, and
Blade~\cite{Guan:2024byi} are also tested as optional IBP reducer. The applicable computer algebra systems include FireFly~\cite{Klappert:2019emp,Klappert:2020aqs} and Fermat~\cite{mc:Fermat}, with Fermatica available as an optional Wolfram interface~\cite{mc:Fermatica}.

For a given sector, the agent can search for candidate integrals with
\texttt{DlogBasis-refine}. It is our adaptation of
\texttt{DlogBasis}~\cite{Henn:2020lye} specially designed for CANON workflow and has been included in it. The agent can also seek a change of basis directly from the differential equations using \texttt{CANONICA}~\cite{Meyer:2016slj,Meyer:2017joq} or \texttt{Libra}~\cite{Lee:2014ioa,Lee:2020zfb}. Table~\ref{mc:provider-table} summarizes how these methods are used in CANON. Once the sectors have been combined, independent routines check whether the full change of basis brings the original equations into the canonical form of Eq.~\eqref{mc:dlog} in every kinematic variable.

\bibliography{references}

@article{Passarino:1978jh,
    author = "Passarino, G. and Veltman, M. J. G.",
    title = "{One Loop Corrections for e+ e- Annihilation Into mu+ mu- in the Weinberg Model}",
    reportNumber = "Print-79-0284 (UTRECHT)",
    doi = "10.1016/0550-3213(79)90234-7",
    journal = "Nucl. Phys. B",
    volume = "160",
    pages = "151--207",
    year = "1979"
}

@article{Ossola:2007ax,
    author = "Ossola, Giovanni and Papadopoulos, Costas G. and Pittau, Roberto",
    title = "{CutTools: A Program implementing the OPP reduction method to compute one-loop amplitudes}",
    eprint = "0711.3596",
    archivePrefix = "arXiv",
    primaryClass = "hep-ph",
    doi = "10.1088/1126-6708/2008/03/042",
    journal = "JHEP",
    volume = "03",
    pages = "042",
    year = "2008"
}

@article{Ossola:2006us,
    author = "Ossola, Giovanni and Papadopoulos, Costas G. and Pittau, Roberto",
    title = "{Reducing full one-loop amplitudes to scalar integrals at the integrand level}",
    eprint = "hep-ph/0609007",
    archivePrefix = "arXiv",
    doi = "10.1016/j.nuclphysb.2006.11.012",
    journal = "Nucl. Phys. B",
    volume = "763",
    pages = "147--169",
    year = "2007"
}

@article{Henn:2013pwa,
    author = "Henn, Johannes M.",
    title = "{Multiloop integrals in dimensional regularization made simple}",
    eprint = "1304.1806",
    archivePrefix = "arXiv",
    primaryClass = "hep-th",
    doi = "10.1103/PhysRevLett.110.251601",
    journal = "Phys. Rev. Lett.",
    volume = "110",
    pages = "251601",
    year = "2013"
}

@article{Henn:2020lye,
    author = "Henn, Johannes and Mistlberger, Bernhard and Smirnov, Vladimir A. and Wasser, Pascal",
    title = "{Constructing d-log integrands and computing master integrals for three-loop four-particle scattering}",
    eprint = "2002.09492",
    archivePrefix = "arXiv",
    primaryClass = "hep-ph",
    reportNumber = "MPP-2020-13, MITP/20-006, MIT-CTP/5176",
    doi = "10.1007/JHEP04(2020)167",
    journal = "JHEP",
    volume = "2020",
    pages = "167",
    year = "2020",
    number = "04"
}

@article{Chicherin:2018old,
    author = "Chicherin, D. and Gehrmann, T. and Henn, J. M. and Wasser, P. and Zhang, Y. and Zoia, S.",
    title = "{All Master Integrals for Three-Jet Production at Next-to-Next-to-Leading Order}",
    eprint = "1812.11160",
    archivePrefix = "arXiv",
    primaryClass = "hep-ph",
    doi = "10.1103/PhysRevLett.123.041603",
    journal = "Phys. Rev. Lett.",
    volume = "123",
    number = "4",
    pages = "041603",
    year = "2019"
}

@article{Meyer:2017joq,
    author = "Meyer, Christoph",
    title = "{Algorithmic transformation of multi-loop master integrals to a canonical basis with CANONICA}",
    eprint = "1705.06252",
    archivePrefix = "arXiv",
    primaryClass = "hep-ph",
    reportNumber = "HU-EP-17-10",
    doi = "10.1016/j.cpc.2017.09.014",
    journal = "Comput. Phys. Commun.",
    volume = "222",
    pages = "295--312",
    year = "2018"
}

@article{Lee:2020zfb,
    author = "Lee, Roman N.",
    title = "{Libra: A package for transformation of differential systems for multiloop integrals}",
    eprint = "2012.00279",
    archivePrefix = "arXiv",
    primaryClass = "hep-ph",
    doi = "10.1016/j.cpc.2021.108058",
    journal = "Comput. Phys. Commun.",
    volume = "267",
    pages = "108058",
    year = "2021"
}

@article{Dlapa:2020cwj,
    author = "Dlapa, Christoph and Henn, Johannes and Yan, Kai",
    title = "{Deriving canonical differential equations for Feynman integrals from a single uniform weight integral}",
    eprint = "2002.02340",
    archivePrefix = "arXiv",
    primaryClass = "hep-ph",
    reportNumber = "MPP-2020-10",
    doi = "10.1007/JHEP05(2020)025",
    journal = "JHEP",
    volume = "2020",
    pages = "025",
    year = "2020",
    number = "05"
}

@article{Meyer:2016slj,
    author = "Meyer, Christoph",
    title = "{Transforming differential equations of multi-loop Feynman integrals into canonical form}",
    eprint = "1611.01087",
    archivePrefix = "arXiv",
    primaryClass = "hep-ph",
    reportNumber = "HU-EP-16-24",
    doi = "10.1007/JHEP04(2017)006",
    journal = "JHEP",
    volume = "2017",
    pages = "006",
    year = "2017",
    number = "04"
}

@article{Lee:2014ioa,
    author = "Lee, Roman N.",
    title = "{Reducing differential equations for multiloop master integrals}",
    eprint = "1411.0911",
    archivePrefix = "arXiv",
    primaryClass = "hep-ph",
    doi = "10.1007/JHEP04(2015)108",
    journal = "JHEP",
    volume = "2015",
    pages = "108",
    year = "2015",
    number = "04"
}

@article{Hoschele:2014qsa,
    author = {H{\"o}schele, Maik and Hoff, Jens and Ueda, Takahiro},
    title = "{Adequate bases of phase space master integrals for gg $\to$ h at NNLO and beyond}",
    eprint = "1407.4049",
    archivePrefix = "arXiv",
    primaryClass = "hep-ph",
    reportNumber = "SFB-CPP-14-38, TTP-14-20, LPN14-087",
    doi = "10.1007/JHEP09(2014)116",
    journal = "JHEP",
    volume = "2014",
    pages = "116",
    year = "2014",
    number = "09"
}

@article{Chetyrkin:1981qh,
    author = "Chetyrkin, K. G. and Tkachov, F. V.",
    title = "{Integration by parts: The algorithm to calculate $\beta$-functions in 4 loops}",
    doi = "10.1016/0550-3213(81)90199-1",
    journal = "Nucl. Phys. B",
    volume = "192",
    pages = "159--204",
    year = "1981"
}

@article{Gehrmann:1999as,
    author = "Gehrmann, T. and Remiddi, E.",
    title = "{Differential equations for two-loop four-point functions}",
    eprint = "hep-ph/9912329",
    archivePrefix = "arXiv",
    reportNumber = "TTP-99-49",
    doi = "10.1016/S0550-3213(00)00223-6",
    journal = "Nucl. Phys. B",
    volume = "580",
    pages = "485--518",
    year = "2000"
}

@article{Dlapa:2021qsl,
    author = "Dlapa, Christoph and Li, Xiaodi and Zhang, Yang",
    title = "{Leading singularities in Baikov representation and Feynman integrals with uniform transcendental weight}",
    eprint = "2103.04638",
    archivePrefix = "arXiv",
    primaryClass = "hep-th",
    reportNumber = "MPP-2021-23, PCFT-21-10, USTC-ICTS-21-10",
    doi = "10.1007/JHEP07(2021)227",
    journal = "JHEP",
    volume = "2021",
    pages = "227",
    year = "2021",
    number = "07"
}

@article{Liu:2024ont,
    author = "Liu, Yuanche and Matija{\v{s}}i{\'c}, Antonela and Miczajka, Julian and Xu, Yingxuan and Xu, Yongqun and Zhang, Yang",
    title = "{Analytic computation of three-loop five-point Feynman integrals}",
    eprint = "2411.18697",
    archivePrefix = "arXiv",
    primaryClass = "hep-ph",
    reportNumber = "HU-EP-24/38-RTG, MITP-24-085, MPP-2024-235, USTC-ICTS/PCFT-24-50",
    doi = "10.1103/qrk2-cym5",
    journal = "Phys. Rev. D",
    volume = "112",
    number = "1",
    pages = "016021",
    year = "2025"
}

@misc{Baikov:1996cd,
    author = "Baikov, P. A.",
    title = "{Explicit solutions of n loop vacuum integral recurrence relations}",
    eprint = "hep-ph/9604254",
    archivePrefix = "arXiv",
    reportNumber = "INP-96-10-418",
    month = "4",
    year = "1996"
}

@article{Frellesvig:2017aai,
    author = "Frellesvig, Hjalte and Papadopoulos, Costas G.",
    title = "{Cuts of Feynman Integrals in Baikov representation}",
    eprint = "1701.07356",
    archivePrefix = "arXiv",
    primaryClass = "hep-ph",
    doi = "10.1007/JHEP04(2017)083",
    journal = "JHEP",
    volume = "2017",
    pages = "083",
    year = "2017",
    number = "04"
}

@article{Chen:2022lzr,
    author = "Chen, Jiaqi and Jiang, Xuhang and Ma, Chichuan and Xu, Xiaofeng and Yang, Li Lin",
    title = "{Baikov representations, intersection theory, and canonical Feynman integrals}",
    eprint = "2202.08127",
    archivePrefix = "arXiv",
    primaryClass = "hep-th",
    doi = "10.1007/JHEP07(2022)066",
    journal = "JHEP",
    volume = "2022",
    pages = "066",
    year = "2022",
    number = "07"
}

@article{He:2022ctv,
    author = "He, Song and Li, Zhenjie and Ma, Rourou and Wu, Zihao and Yang, Qinglin and Zhang, Yang",
    title = "{A study of Feynman integrals with uniform transcendental weights and their symbology}",
    eprint = "2206.04609",
    archivePrefix = "arXiv",
    primaryClass = "hep-th",
    reportNumber = "USTC-ICTS/PCFT-22-17",
    doi = "10.1007/JHEP10(2022)165",
    journal = "JHEP",
    volume = "2022",
    pages = "165",
    year = "2022",
    number = "10"
}

@article{Henn:2024ngj,
    author = "Henn, Johannes M. and Matija{\v{s}}i{\'c}, Antonela and Miczajka, Julian and Peraro, Tiziano and Xu, Yingxuan and Zhang, Yang",
    title = "{A computation of two-loop six-point Feynman integrals in dimensional regularization}",
    eprint = "2403.19742",
    archivePrefix = "arXiv",
    primaryClass = "hep-ph",
    reportNumber = "MPP-2024-53, USTC-ICTS/PCFT-24-11",
    doi = "10.1007/JHEP08(2024)027",
    journal = "JHEP",
    volume = "2024",
    pages = "027",
    year = "2024",
    number = "08"
}

@article{Laporta:2000dsw,
    author = "Laporta, S.",
    title = "{High-precision calculation of multiloop Feynman integrals by difference equations}",
    eprint = "hep-ph/0102033",
    archivePrefix = "arXiv",
    doi = "10.1142/S0217751X00002159",
    journal = "Int. J. Mod. Phys. A",
    volume = "15",
    pages = "5087--5159",
    year = "2000"
}

@article{Maierhofer:2017gsa,
    author = {Maierh{\"o}fer, Philipp and Usovitsch, Johann and Uwer, Peter},
    title = "{Kira{\textemdash}A Feynman integral reduction program}",
    eprint = "1705.05610",
    archivePrefix = "arXiv",
    primaryClass = "hep-ph",
    doi = "10.1016/j.cpc.2018.04.012",
    journal = "Comput. Phys. Commun.",
    volume = "230",
    pages = "99--112",
    year = "2018"
}

@article{Klappert:2020nbg,
    author = {Klappert, Jonas and Lange, Fabian and Maierh{\"o}fer, Philipp and Usovitsch, Johann},
    title = "{Integral reduction with Kira 2.0 and finite field methods}",
    eprint = "2008.06494",
    archivePrefix = "arXiv",
    primaryClass = "hep-ph",
    reportNumber = "TTK-20-24, P3H-20-041, FR-PHENO-2020-11, MITP/20-044",
    doi = "10.1016/j.cpc.2021.108024",
    journal = "Comput. Phys. Commun.",
    volume = "266",
    pages = "108024",
    year = "2021"
}

@article{Lange:2025fba,
    author = "Lange, Fabian and Usovitsch, Johann and Wu, Zihao",
    title = "{Kira 3: integral reduction with efficient seeding and optimized equation selection}",
    eprint = "2505.20197",
    archivePrefix = "arXiv",
    primaryClass = "hep-ph",
    reportNumber = "ZU-TH 39/25, HU-EP-25/17-RTG",
    doi = "10.1016/j.cpc.2025.109999",
    journal = "Comput. Phys. Commun.",
    volume = "322",
    pages = "109999",
    year = "2026"
}

@article{Smirnov:2025prc,
    author = "Smirnov, Alexander V. and Zeng, Mao",
    title = "{FIRE 7: Automatic Reduction with Modular Approach}",
    eprint = "2510.07150",
    archivePrefix = "arXiv",
    primaryClass = "hep-ph",
    doi = "10.1016/j.cpc.2026.110200",
    journal = "Comput. Phys. Commun.",
    volume = "325",
    pages = "110200",
    year = "2026"
}

@article{Wu:2023upw,
    author = "Wu, Zihao and Boehm, Janko and Ma, Rourou and Xu, Hefeng and Zhang, Yang",
    title = "{NeatIBP 1.0, a package generating small-size integration-by-parts relations for Feynman integrals}",
    eprint = "2305.08783",
    archivePrefix = "arXiv",
    primaryClass = "hep-ph",
    reportNumber = "USTC-ICTS/PCFT-23-15",
    doi = "10.1016/j.cpc.2023.108999",
    journal = "Comput. Phys. Commun.",
    volume = "295",
    pages = "108999",
    year = "2024"
}

@article{Wu:2025aeg,
    author = {Wu, Zihao and B{\"o}hm, Janko and Ma, Rourou and Usovitsch, Johann and Xu, Yingxuan and Zhang, Yang},
    title = "{Performing integration-by-parts reductions using NeatIBP 1.1 + Kira}",
    eprint = "2502.20778",
    archivePrefix = "arXiv",
    primaryClass = "hep-ph",
    reportNumber = "USTC-ICTS/PCFT-25-10, MPP-2025-29, HU-EP-25/12-RTG",
    doi = "10.1016/j.cpc.2025.109798",
    journal = "Comput. Phys. Commun.",
    volume = "316",
    pages = "109798",
    year = "2025"
}

@article{Guan:2024byi,
    author = "Guan, Xin and Liu, Xiao and Ma, Yan-Qing and Wu, Wen-Hao",
    title = "{Blade: A package for block-triangular form improved Feynman integrals decomposition}",
    eprint = "2405.14621",
    archivePrefix = "arXiv",
    primaryClass = "hep-ph",
    doi = "10.1016/j.cpc.2025.109538",
    journal = "Comput. Phys. Commun.",
    volume = "310",
    pages = "109538",
    year = "2025"
}

@article{Klappert:2019emp,
    author = "Klappert, Jonas and Lange, Fabian",
    title = "{Reconstructing rational functions with FireFly}",
    eprint = "1904.00009",
    archivePrefix = "arXiv",
    primaryClass = "cs.SC",
    reportNumber = "TTK-19-11, P3H-19-007",
    doi = "10.1016/j.cpc.2019.106951",
    journal = "Comput. Phys. Commun.",
    volume = "247",
    pages = "106951",
    year = "2020"
}

@article{Klappert:2020aqs,
    author = "Klappert, Jonas and Klein, Sven Yannick and Lange, Fabian",
    title = "{Interpolation of dense and sparse rational functions and other improvements in FireFly}",
    eprint = "2004.01463",
    archivePrefix = "arXiv",
    primaryClass = "cs.MS",
    reportNumber = "TTK-20-07, P3H-20-010",
    doi = "10.1016/j.cpc.2021.107968",
    journal = "Comput. Phys. Commun.",
    volume = "264",
    pages = "107968",
    year = "2021"
}

@misc{mc:Fermat,
    author = {Lewis, Robert H.},
    title = {{Fermat}: Computer Algebra System},
    howpublished = {\url{https://home.bway.net/lewis/}},
    note = {Accessed 8 October 2026}
}

@misc{mc:Fermatica,
    author = {Lee, Roman N.},
    title = {{Fermatica}},
    howpublished = {\url{https://github.com/rnlg/Fermatica}},
    note = {Wolfram interface to Fermat; accessed 8 October 2026}
}

@article{Gehrmann:2018yef,
    author = "Gehrmann, T. and Henn, J. M. and Lo Presti, N. A.",
    title = "{Pentagon functions for massless planar scattering amplitudes}",
    eprint = "1807.09812",
    archivePrefix = "arXiv",
    primaryClass = "hep-ph",
    reportNumber = "ZU-TH-25-18, ZU-TH 25/18, MPP-2018-180, MITP-18-067, IPPP-18-59, MITP/18-067, IPPP/18/59",
    doi = "10.1007/JHEP10(2018)103",
    journal = "JHEP",
    volume = "2018",
    pages = "103",
    year = "2018",
    number = "10"
}

@article{Azzi:2019yne,
    author = "Azzi, P. and others",
    editor = "Dainese, Andrea and Mangano, Michelangelo and Meyer, Andreas B. and Nisati, Aleandro and Salam, Gavin and Vesterinen, Mika Anton",
    title = "{Report from Working Group 1}: {Standard Model Physics at the HL-LHC and HE-LHC}",
    eprint = "1902.04070",
    archivePrefix = "arXiv",
    primaryClass = "hep-ph",
    reportNumber = "CERN-LPCC-2018-03",
    doi = "10.23731/CYRM-2019-007.1",
    journal = "CERN Yellow Rep. Monogr.",
    volume = "7",
    pages = "1--220",
    year = "2019"
}

@article{Heinrich:2020ybq,
    author = "Heinrich, Gudrun",
    title = "{Collider Physics at the Precision Frontier}",
    eprint = "2009.00516",
    archivePrefix = "arXiv",
    primaryClass = "hep-ph",
    reportNumber = "KA-TP-13-2020, P3H-20-044",
    doi = "10.1016/j.physrep.2021.03.006",
    journal = "Phys. Rept.",
    volume = "922",
    pages = "1--69",
    year = "2021"
}

@article{Huss:2025nlt,
    author = {Huss, Alexander and Huston, Joey and Jones, Stephen and Pellen, Mathieu and R{\"o}ntsch, Raoul},
    title = "{Les Houches 2023 {\textendash} physics at TeV colliders: Report on the Standard Model precision wishlist}",
    eprint = "2504.06689",
    archivePrefix = "arXiv",
    primaryClass = "hep-ph",
    reportNumber = "CERN-TH-2025-071, FR-PHENO-2025-005, IPPP/25/19, TIF-UNIMI-2025-9",
    doi = "10.21468/SciPostPhysCommRep.21",
    year = "2026",
    journal = "SciPost Phys. Comm. Rep.",
    pages = "21"
}

@book{Weinzierl:2022eaz,
    author = "Weinzierl, Stefan",
    title = "{Feynman Integrals. A Comprehensive Treatment for Students and Researchers}",
    eprint = "2201.03593",
    archivePrefix = "arXiv",
    primaryClass = "hep-th",
    reportNumber = "MITP/22-001",
    doi = "10.1007/978-3-030-99558-4",
    isbn = "978-3-030-99557-7, 978-3-030-99560-7, 978-3-030-99558-4",
    publisher = "Springer",
    series = "UNITEXT for Physics",
    year = "2022"
}

@article{Abreu:2022mfk,
    author = "Abreu, Samuel and Britto, Ruth and Duhr, Claude",
    title = "{The SAGEX review on scattering amplitudes Chapter 3: Mathematical structures in Feynman integrals}",
    eprint = "2203.13014",
    archivePrefix = "arXiv",
    primaryClass = "hep-th",
    reportNumber = "SAGEX-22-04, BONN-TH-2022-03, CERN-TH-2022-021",
    doi = "10.1088/1751-8121/ac87de",
    journal = "J. Phys. A",
    volume = "55",
    number = "44",
    pages = "443004",
    year = "2022"
}

@article{Henn:2014qga,
    author = "Henn, Johannes M.",
    title = "{Lectures on differential equations for Feynman integrals}",
    eprint = "1412.2296",
    archivePrefix = "arXiv",
    primaryClass = "hep-ph",
    doi = "10.1088/1751-8113/48/15/153001",
    journal = "J. Phys. A",
    volume = "48",
    pages = "153001",
    year = "2015"
}

@article{Remiddi:1997ny,
    author = "Remiddi, Ettore",
    title = "{Differential equations for Feynman graph amplitudes}",
    eprint = "hep-th/9711188",
    archivePrefix = "arXiv",
    reportNumber = "DFUB-97-15, DFUB 97-15",
    doi = "10.1007/BF03185566",
    journal = "Nuovo Cim. A",
    volume = "110",
    pages = "1435--1452",
    year = "1997"
}

@article{Arkani-Hamed:2014via,
    author = "Arkani-Hamed, Nima and Bourjaily, Jacob L. and Cachazo, Freddy and Trnka, Jaroslav",
    title = "{Singularity Structure of Maximally Supersymmetric Scattering Amplitudes}",
    eprint = "1410.0354",
    archivePrefix = "arXiv",
    primaryClass = "hep-th",
    reportNumber = "CALT-TH-2014-154",
    doi = "10.1103/PhysRevLett.113.261603",
    journal = "Phys. Rev. Lett.",
    volume = "113",
    number = "26",
    pages = "261603",
    year = "2014"
}

@article{Bern:2014kca,
    author = "Bern, Zvi and Herrmann, Enrico and Litsey, Sean and Stankowicz, James and Trnka, Jaroslav",
    title = "{Logarithmic Singularities and Maximally Supersymmetric Amplitudes}",
    eprint = "1412.8584",
    archivePrefix = "arXiv",
    primaryClass = "hep-th",
    reportNumber = "CALT-TH-2014-172, UCLA-14-TEP-109",
    doi = "10.1007/JHEP06(2015)202",
    journal = "JHEP",
    volume = "2015",
    pages = "202",
    year = "2015",
    number = "06"
}

@article{Herrmann:2019upk,
    author = "Herrmann, Enrico and Parra-Martinez, Julio",
    title = "{Logarithmic forms and differential equations for Feynman integrals}",
    eprint = "1909.04777",
    archivePrefix = "arXiv",
    primaryClass = "hep-th",
    doi = "10.1007/JHEP02(2020)099",
    journal = "JHEP",
    volume = "2020",
    pages = "099",
    year = "2020",
    number = "02"
}

@article{Tarasov:1996br,
    author = "Tarasov, O. V.",
    title = "{Connection between Feynman integrals having different values of the space-time dimension}",
    eprint = "hep-th/9606018",
    archivePrefix = "arXiv",
    reportNumber = "DESY-96-068, JINR-E2-96-62",
    doi = "10.1103/PhysRevD.54.6479",
    journal = "Phys. Rev. D",
    volume = "54",
    pages = "6479--6490",
    year = "1996"
}

@article{vonManteuffel:2014qoa,
    author = "von Manteuffel, Andreas and Panzer, Erik and Schabinger, Robert M.",
    title = "{A quasi-finite basis for multi-loop Feynman integrals}",
    eprint = "1411.7392",
    archivePrefix = "arXiv",
    primaryClass = "hep-ph",
    reportNumber = "MITP-14-076",
    doi = "10.1007/JHEP02(2015)120",
    journal = "JHEP",
    volume = "2015",
    pages = "120",
    year = "2015",
    number = "02"
}

@article{Schabinger:2018dyi,
    author = "Schabinger, Robert M.",
    title = "{Constructing multiloop scattering amplitudes with manifest singularity structure}",
    eprint = "1806.05682",
    archivePrefix = "arXiv",
    primaryClass = "hep-th",
    reportNumber = "MSUHEP-18-009",
    doi = "10.1103/PhysRevD.99.105010",
    journal = "Phys. Rev. D",
    volume = "99",
    number = "10",
    pages = "105010",
    year = "2019"
}

@article{Mastrolia:2018uzb,
    author = "Mastrolia, Pierpaolo and Mizera, Sebastian",
    title = "{Feynman Integrals and Intersection Theory}",
    eprint = "1810.03818",
    archivePrefix = "arXiv",
    primaryClass = "hep-th",
    doi = "10.1007/JHEP02(2019)139",
    journal = "JHEP",
    volume = "2019",
    pages = "139",
    year = "2019",
    number = "02"
}

@article{Chen:2020uyk,
    author = "Chen, Jiaqi and Jiang, Xuhang and Xu, Xiaofeng and Yang, Li Lin",
    title = "{Constructing canonical Feynman integrals with intersection theory}",
    eprint = "2008.03045",
    archivePrefix = "arXiv",
    primaryClass = "hep-th",
    doi = "10.1016/j.physletb.2021.136085",
    journal = "Phys. Lett. B",
    volume = "814",
    pages = "136085",
    year = "2021"
}

@article{Chen:2023kgw,
    author = "Chen, Jiaqi and Feng, Bo and Yang, Li Lin",
    title = "{Intersection theory rules symbology}",
    eprint = "2305.01283",
    archivePrefix = "arXiv",
    primaryClass = "hep-th",
    doi = "10.1007/s11433-023-2239-8",
    journal = "Sci. China Phys. Mech. Astron.",
    volume = "67",
    number = "2",
    pages = "221011",
    year = "2024"
}

@article{He:2023umf,
    author = "He, Song and Jiang, Xuhang and Liu, Jiahao and Yang, Qinglin",
    title = "{On symbology and differential equations of Feynman integrals from Schubert analysis}",
    eprint = "2309.16441",
    archivePrefix = "arXiv",
    primaryClass = "hep-th",
    doi = "10.1007/JHEP12(2023)140",
    journal = "JHEP",
    volume = "2023",
    pages = "140",
    year = "2023",
    note = "[Erratum: JHEP 04, 063 (2024)]",
    number = "12"
}

@article{Duhr:2024xsy,
    author = "Duhr, Claude and Porkert, Franziska and Semper, Cathrin and Stawinski, Sven F.",
    title = "{Self-duality from twisted cohomology}",
    eprint = "2408.04904",
    archivePrefix = "arXiv",
    primaryClass = "hep-th",
    reportNumber = "BONN-TH-2024-11",
    doi = "10.1007/JHEP03(2025)053",
    journal = "JHEP",
    volume = "2025",
    pages = "053",
    year = "2025",
    number = "03"
}

@article{Duhr:2025xyy,
    author = "Duhr, Claude and Maggio, Sara and Porkert, Franziska and Semper, Cathrin and Sohnle, Yoann and Stawinski, Sven F.",
    title = "{Canonical differential equations and intersection matrices}",
    eprint = "2509.17787",
    archivePrefix = "arXiv",
    primaryClass = "hep-th",
    reportNumber = "BONN-TH/2025-30, UUITP--27/25",
    doi = "10.1007/JHEP02(2026)211",
    journal = "JHEP",
    volume = "2026",
    pages = "211",
    year = "2026",
    number = "02"
}

@article{Chaubey:2025adn,
    author = "Chaubey, Ekta and Sotnikov, Vasily",
    title = "{Elliptic Leading Singularities and Canonical Integrands}",
    eprint = "2504.20897",
    archivePrefix = "arXiv",
    primaryClass = "hep-th",
    reportNumber = "BONN-TH-2025-18, ZU-TH 29/25",
    doi = "10.1103/4fjc-lfnx",
    journal = "Phys. Rev. Lett.",
    volume = "135",
    number = "10",
    pages = "101903",
    year = "2025"
}

@article{Forner:2026vby,
    author = "Forner, Felix and Mella, Cesare Carlo and Nega, Christoph and Tancredi, Lorenzo and Wagner, Fabian J.",
    title = "{Integrand analysis, leading singularities and canonical bases beyond polylogarithms}",
    eprint = "2604.25270",
    archivePrefix = "arXiv",
    primaryClass = "hep-th",
    reportNumber = "OUTP-26-03P, TUM-HEP-1601/26",
    doi = "10.1007/JHEP10(2026)213",
    journal = "JHEP",
    volume = "2026",
    pages = "213",
    year = "2026",
    number = "10"
}

@article{Tkachov:1981wb,
    author = "Tkachov, F. V.",
    title = "{A theorem on analytical calculability of 4-loop renormalization group functions}",
    doi = "10.1016/0370-2693(81)90288-4",
    journal = "Phys. Lett. B",
    volume = "100",
    pages = "65--68",
    year = "1981"
}

@article{Kotikov:1990kg,
    author = "Kotikov, A. V.",
    title = "{Differential equations method: New technique for massive Feynman diagrams calculation}",
    reportNumber = "ITF-90-31E",
    doi = "10.1016/0370-2693(91)90413-K",
    journal = "Phys. Lett. B",
    volume = "254",
    pages = "158--164",
    year = "1991"
}

@misc{Lee:2017oca,
    author = "Lee, Roman N. and Pomeransky, Andrei A.",
    title = "{Normalized Fuchsian form on Riemann sphere and differential equations for multiloop integrals}",
    eprint = "1707.07856",
    archivePrefix = "arXiv",
    primaryClass = "hep-th",
    month = "7",
    year = "2017"
}

@article{Prausa:2017ltv,
    author = "Prausa, Mario",
    title = "{epsilon: A tool to find a canonical basis of master integrals}",
    eprint = "1701.00725",
    archivePrefix = "arXiv",
    primaryClass = "hep-ph",
    reportNumber = "TTK-17-01",
    doi = "10.1016/j.cpc.2017.05.026",
    journal = "Comput. Phys. Commun.",
    volume = "219",
    pages = "361--376",
    year = "2017"
}

@article{Gituliar:2017vzm,
    author = "Gituliar, Oleksandr and Magerya, Vitaly",
    title = "{Fuchsia: a tool for reducing differential equations for Feynman master integrals to epsilon form}",
    eprint = "1701.04269",
    archivePrefix = "arXiv",
    primaryClass = "hep-ph",
    reportNumber = "DESY-16-219",
    doi = "10.1016/j.cpc.2017.05.004",
    journal = "Comput. Phys. Commun.",
    volume = "219",
    pages = "329--338",
    year = "2017"
}

@article{Argeri:2014qva,
    author = "Argeri, Mario and Di Vita, Stefano and Mastrolia, Pierpaolo and Mirabella, Edoardo and Schlenk, Johannes and Schubert, Ulrich and Tancredi, Lorenzo",
    title = "{Magnus and Dyson Series for Master Integrals}",
    eprint = "1401.2979",
    archivePrefix = "arXiv",
    primaryClass = "hep-ph",
    doi = "10.1007/JHEP03(2014)082",
    journal = "JHEP",
    volume = "2014",
    pages = "082",
    year = "2014",
    number = "03"
}

@article{Gehrmann:2014bfa,
    author = "Gehrmann, Thomas and von Manteuffel, Andreas and Tancredi, Lorenzo and Weihs, Erich",
    title = "{The two-loop master integrals for $q\overline{q} \to VV$}",
    eprint = "1404.4853",
    archivePrefix = "arXiv",
    primaryClass = "hep-ph",
    reportNumber = "ZU-TH-16-14, LPN14-065-, MITP-14-021, LPN14-065",
    doi = "10.1007/JHEP06(2014)032",
    journal = "JHEP",
    volume = "2014",
    pages = "032",
    year = "2014",
    number = "06"
}

@article{Adams:2018yfj,
    author = "Adams, Luise and Weinzierl, Stefan",
    title = "{The $\varepsilon$-form of the differential equations for Feynman integrals in the elliptic case}",
    eprint = "1802.05020",
    archivePrefix = "arXiv",
    primaryClass = "hep-ph",
    reportNumber = "MITP-18-011",
    doi = "10.1016/j.physletb.2018.04.002",
    journal = "Phys. Lett. B",
    volume = "781",
    pages = "270--278",
    year = "2018"
}

@article{Frellesvig:2021hkr,
    author = "Frellesvig, Hjalte",
    title = "{On epsilon factorized differential equations for elliptic Feynman integrals}",
    eprint = "2110.07968",
    archivePrefix = "arXiv",
    primaryClass = "hep-th",
    doi = "10.1007/JHEP03(2022)079",
    journal = "JHEP",
    volume = "2022",
    pages = "079",
    year = "2022",
    number = "03"
}

@article{Dlapa:2022wdu,
    author = "Dlapa, Christoph and Henn, Johannes M. and Wagner, Fabian J.",
    title = "{An algorithmic approach to finding canonical differential equations for elliptic Feynman integrals}",
    eprint = "2211.16357",
    archivePrefix = "arXiv",
    primaryClass = "hep-ph",
    reportNumber = "DESY 22-189, MPP-2022-138, TUM-HEP-1427/22",
    doi = "10.1007/JHEP08(2023)120",
    journal = "JHEP",
    volume = "2023",
    pages = "120",
    year = "2023",
    number = "08"
}

@article{Gorges:2023zgv,
    author = {G{\"o}rges, Lennard and Nega, Christoph and Tancredi, Lorenzo and Wagner, Fabian J.},
    title = "{On a procedure to derive {\ensuremath{\epsilon}}-factorised differential equations beyond polylogarithms}",
    eprint = "2305.14090",
    archivePrefix = "arXiv",
    primaryClass = "hep-th",
    doi = "10.1007/JHEP07(2023)206",
    journal = "JHEP",
    volume = "2023",
    pages = "206",
    year = "2023",
    number = "07"
}

@article{Frellesvig:2023iwr,
    author = "Frellesvig, Hjalte and Weinzierl, Stefan",
    title = "{On $\varepsilon$-factorised bases and pure Feynman integrals}",
    eprint = "2301.02264",
    archivePrefix = "arXiv",
    primaryClass = "hep-th",
    reportNumber = "MITP/23-001",
    doi = "10.21468/SciPostPhys.16.6.150",
    journal = "SciPost Phys.",
    volume = "16",
    number = "6",
    pages = "150",
    year = "2024"
}

@article{Pogel:2022vat,
    author = {P{\"o}gel, Sebastian and Wang, Xing and Weinzierl, Stefan},
    title = "{Bananas of equal mass: any loop, any order in the dimensional regularisation parameter}",
    eprint = "2212.08908",
    archivePrefix = "arXiv",
    primaryClass = "hep-th",
    doi = "10.1007/JHEP04(2023)117",
    journal = "JHEP",
    volume = "2023",
    pages = "117",
    year = "2023",
    number = "04"
}

@article{Duhr:2025lbz,
    author = "Duhr, Claude and Maggio, Sara and Nega, Christoph and Sauer, Benjamin and Tancredi, Lorenzo and Wagner, Fabian J.",
    title = "{Aspects of canonical differential equations for Calabi-Yau geometries and beyond}",
    eprint = "2503.20655",
    archivePrefix = "arXiv",
    primaryClass = "hep-th",
    reportNumber = "BONN-TH-2025-11, TUM-HEP 1559/25, HU-EP-25/13-RTG",
    doi = "10.1007/JHEP06(2025)128",
    journal = "JHEP",
    volume = "2025",
    pages = "128",
    year = "2025",
    number = "06"
}

@article{Maggio:2025jel,
    author = "Maggio, Sara and Sohnle, Yoann",
    title = "{On canonical differential equations for Calabi-Yau multi-scale Feynman integrals}",
    eprint = "2504.17757",
    archivePrefix = "arXiv",
    primaryClass = "hep-th",
    reportNumber = "BONN-TH-2025-16, UUITP--14/25",
    doi = "10.1007/JHEP10(2025)202",
    journal = "JHEP",
    volume = "2025",
    pages = "202",
    year = "2025",
    number = "10"
}

@article{Bree:2025tug,
    author = "Bree, Iris and others",
    collaboration = "{\ensuremath{\epsilon}}",
    title = "{New algorithms for Feynman integral reduction and epsilon-factorized differential equations}",
    eprint = "2511.15381",
    archivePrefix = "arXiv",
    primaryClass = "hep-th",
    doi = "10.1103/mjpn-61yv",
    journal = "Phys. Rev. D",
    volume = "113",
    number = "11",
    pages = "116019",
    year = "2026"
}

@article{e-collaboration:2025frv,
    author = "Bree, Iris and others",
    collaboration = "{\ensuremath{\varepsilon}}-collaboration",
    title = "{Geometric Bookkeeping Guide to Feynman Integral Reduction and {\ensuremath{\epsilon}}-Factorized Differential Equations}",
    eprint = "2506.09124",
    archivePrefix = "arXiv",
    primaryClass = "hep-th",
    doi = "10.1103/pyt8-d7rt",
    journal = "Phys. Rev. Lett.",
    volume = "136",
    number = "24",
    pages = "241602",
    year = "2026"
}

@misc{Jimenez-Santacruz:2026ecm,
    author = "Jimenez-Santacruz, Mateo and Lopez-Arcos, Cristhiam and Quintero Velez, Alexander",
    title = "{Canonical differential equations for Feynman integrals from $\mathcal{A}$-hypergeometric systems in the Schwinger representation}",
    eprint = "2609.16107",
    archivePrefix = "arXiv",
    primaryClass = "hep-th",
    month = "9",
    year = "2026"
}

@article{Arkani-Hamed:2010pyv,
    author = "Arkani-Hamed, Nima and Bourjaily, Jacob L. and Cachazo, Freddy and Trnka, Jaroslav",
    title = "{Local Integrals for Planar Scattering Amplitudes}",
    eprint = "1012.6032",
    archivePrefix = "arXiv",
    primaryClass = "hep-th",
    doi = "10.1007/JHEP06(2012)125",
    journal = "JHEP",
    volume = "2012",
    pages = "125",
    year = "2012",
    number = "06"
}

@article{Chen:2025hzq,
    author = "Chen, Jiaqi and Yang, Li Lin and Zhang, Yiyang",
    title = "{On an approach to canonicalizing elliptic Feynman integrals}",
    eprint = "2503.23720",
    archivePrefix = "arXiv",
    primaryClass = "hep-th",
    doi = "10.1007/JHEP04(2026)077",
    journal = "JHEP",
    volume = "2026",
    pages = "077",
    year = "2026",
    number = "04"
}

@misc{Yang:2025ofz,
    author = "Yang, Li Lin and Zhang, Yiyang",
    title = "{From $\mathrm{d} \! \log$ to $\mathrm{d} \mathcal{E}$: Canonical Elliptic Integrands and Modular Symbol Letters with Pure eMPLs}",
    eprint = "2512.19370",
    archivePrefix = "arXiv",
    primaryClass = "hep-th",
    month = "12",
    year = "2025"
}

@article{Badger:2023mgf,
    author = "Badger, Simon and Czakon, Micha{\l} and Hartanto, Heribertus Bayu and Moodie, Ryan and Peraro, Tiziano and Poncelet, Rene and Zoia, Simone",
    title = "{Isolated photon production in association with a jet pair through next-to-next-to-leading order in QCD}",
    eprint = "2304.06682",
    archivePrefix = "arXiv",
    primaryClass = "hep-ph",
    reportNumber = "Cavendish-HEP-23/02, P3H-23-022, TTK-23-09",
    doi = "10.1007/JHEP10(2023)071",
    journal = "JHEP",
    volume = "2023",
    pages = "071",
    year = "2023",
    number = "10"
}

@article{Agarwal:2023suw,
    author = "Agarwal, Bakul and Buccioni, Federico and Devoto, Federica and Gambuti, Giulio and von Manteuffel, Andreas and Tancredi, Lorenzo",
    title = "{Five-parton scattering in QCD at two loops}",
    eprint = "2311.09870",
    archivePrefix = "arXiv",
    primaryClass = "hep-ph",
    reportNumber = "KA-TP-27-2023, MSUHEP-23-030, OUTP-23-12P, P3H-23-088,
  TUM-HEP-1481/23",
    doi = "10.1103/PhysRevD.109.094025",
    journal = "Phys. Rev. D",
    volume = "109",
    number = "9",
    pages = "094025",
    year = "2024"
}

@article{Becchetti:2025rrz,
    author = "Becchetti, Matteo and Coro, Federico and Nega, Christoph and Tancredi, Lorenzo and Wagner, Fabian J.",
    title = "{Analytic two-loop amplitudes for $ q\overline{q}\to \gamma \gamma $ and gg {\textrightarrow} {\ensuremath{\gamma}}{\ensuremath{\gamma}} mediated by a heavy-quark loop}",
    eprint = "2502.00118",
    archivePrefix = "arXiv",
    primaryClass = "hep-ph",
    reportNumber = "TUM-HEP 1554/25",
    doi = "10.1007/JHEP06(2025)033",
    journal = "JHEP",
    volume = "2025",
    pages = "033",
    year = "2025",
    number = "06"
}

@article{Badger:2025ljy,
    author = "Badger, Simon and Becchetti, Matteo and Brancaccio, Colomba and Czakon, Micha{\l} and Hartanto, Heribertus Bayu and Poncelet, Rene and Zoia, Simone",
    title = "{Double virtual QCD corrections to $ t\overline{t} $+jet production at the LHC}",
    eprint = "2511.11424",
    archivePrefix = "arXiv",
    primaryClass = "hep-ph",
    reportNumber = "P3H-25-092, TTK-25-38, ZU-TH 74/25, IFJPAN-IV-2025-23",
    doi = "10.1007/JHEP05(2026)044",
    journal = "JHEP",
    volume = "2026",
    pages = "044",
    year = "2026",
    number = "05"
}

@article{Bi:2023bnq,
    author = "Bi, Huan-Yu and Huang, Li-Hong and Huang, Rui-Jun and Ma, Yan-Qing and Yu, Huai-Min",
    title = "{Electroweak Corrections to Double Higgs Production at the LHC}",
    eprint = "2311.16963",
    archivePrefix = "arXiv",
    primaryClass = "hep-ph",
    doi = "10.1103/PhysRevLett.132.231802",
    journal = "Phys. Rev. Lett.",
    volume = "132",
    number = "23",
    pages = "231802",
    year = "2024"
}

@article{He:2025hin,
    author = "He, Chuan-Qi and Xing, Hongxi and Yang, Tong-Zhi and Zhu, Hua Xing",
    title = "{Single-Inclusive Hadron Production in Electron-Positron Annihilation at Next-to-Next-to-Next-to-Leading Order in QCD}",
    eprint = "2503.20441",
    archivePrefix = "arXiv",
    primaryClass = "hep-ph",
    reportNumber = "ZU-TH 21/25",
    doi = "10.1103/vtz2-6bnm",
    journal = "Phys. Rev. Lett.",
    volume = "135",
    number = "10",
    pages = "101901",
    year = "2025"
}

@article{Chen:2022lwc,
    author = "Chen, Xuan and Gehrmann, Thomas and Glover, Nigel and Huss, Alexander and Yang, Tong-Zhi and Zhu, Hua Xing",
    title = "{Transverse mass distribution and charge asymmetry in W boson production to third order in QCD}",
    eprint = "2205.11426",
    archivePrefix = "arXiv",
    primaryClass = "hep-ph",
    reportNumber = "IPPP/22/32, ZU-TH 19/22, P3H-22-055, KA-TP-15-2022, CERN-TH-2022-083",
    doi = "10.1016/j.physletb.2023.137876",
    journal = "Phys. Lett. B",
    volume = "840",
    pages = "137876",
    year = "2023"
}
\end{document}